\documentstyle[11pt,aaspp4]{article}
\begin{document}
%
%
\title{The Relative Stability Against Merger of Close, Compact Binaries} 
\author{Kimberly C. B. New\altaffilmark{1} and Joel E. Tohline}
\altaffiltext{1}{Currently at the Department of Physics \&
Atmospheric Science, Drexel University, Philadelphia, PA 19104}
\affil{Department of Physics \& Astronomy, Louisiana State University, \\
       Baton Rouge, LA  70803-4001}
%
%
%
\begin{abstract}
The orbital separation of compact binary stars will shrink
with time due to the emission of gravitational radiation.
This inspiralling phase of a binary system's evolution generally
will be very long compared to the system's orbital period, but the
final coalescence may be dynamical and driven to a large degree by 
hydrodynamic effects, particularly if there is a critical 
separation at which the system becomes dynamically unstable 
toward merger.  Indeed, if weakly relativistic systems (such 
as white dwarf--white dwarf binaries) encounter a point of 
dynamical instability at some critically close separation, 
coalescence may be entirely a classical, hydrodynamic process.
Therefore, a proper investigation of this stage of binary
evolution must include three--dimensional hydrodynamic
simulations. 
 
We have constructed equilibrium sequences of synchronously
rotating, equal--mass binaries in circular orbit with a single
parameter -- the binary separation -- varying along each
sequence.  Sequences have been constructed with various
polytropic as well as realistic white dwarf and neutron star 
equations of state.  Using a Newtonian, finite--difference
hydrodynamics code, we have examined the dynamical stability of individual
models along these equilibrium sequences.  Our simulations 
indicate that no points of instability exist on the sequences 
we analyzed that had relatively soft equations of state 
(polytropic sequences with polytropic index $n=1.0$ and $1.5$ 
and two white dwarf sequences).  However, we did identify 
dynamically unstable binary models on sequences with stiffer 
equations of state ($n=0.5$ polytropic sequence and two neutron 
star sequences).  We thus infer that binary systems
with soft equations of state are not driven to merger by a
dynamical instability.  
For the $n=0.5$ polytropic sequence, the separation at
which a dynamical instability sets in appears to be associated
with the minimum energy and angular momentum configuration along
the sequence.  Our simulations suggest but do not conclusively
demonstrate that, in the absence of relativistic effects, this 
same association may also hold for binary neutron star systems.

\end{abstract}
     
%
\section{Introduction}
The coalescence of double white dwarf and double neutron star binaries
is important to examine since this process
may produce a number of astrophysically interesting
objects and events.  Double white dwarf binary mergers
have been suggested as precursors
to some Type Ia Supernovae (Iben \& Tutukov 1984; Iben 1988, 1991;
Iben \& Webbink 1989; Branch et al.\ 1995) and to long 
gamma--ray bursts (Katz \& Canel 1996). White dwarf--white dwarf
mergers may also lead to the formation of
massive single white dwarfs or neutron stars (Colgate \& Petschek
1982; Saio \& Nomoto 1985; Iben \& Tutukov 1986; Kawai, Saio, \& Nomoto 1987;
Chen \& Leonard 1993); to the formation of subdwarf stars; or to the
formation of hydrogen deficient, highly
luminous stars (Iben 1990 and references therein; Webbink 1984).  
Neutron star--neutron star mergers may result in bursts
of gamma--rays and neutrinos, in the production of r--process elements,
and in the formation of black holes (Eichler et al.\ 1989; Meyer 1989;
Narayan, Pacz\'nski, \& Piran 1992; Rasio \& Shapiro 1992; Davies, Benz,
Piran, \& Thielemann 1994; Katz
\& Canel 1996; Lipunov et al.\ 1995; Ruffert, Janka, \& Sch\"afer 1996;
but see the simulations
of Shibata, Nakamura, and Oohara [1993] and Janka \& Ruffert [1996] which 
cast doubt on the neutron star--neutron star
merger scenario as a precursor to gamma--ray bursts).  

Merging compact binaries are also expected to be relatively strong
sources of gravitational radiation.  The gravitational radiation emitted
during the inspiral phase of double neutron star binary evolution
(i.e.{}, before tidal effects become sizeable) is likely to be detected
by terrestrial interferometric detectors such as LIGO and VIRGO, which
will be sensitive to frequencies in the range of 10--$10^3$ Hz
(Abramovici et al.\ 1992; Cutler et al.\ 1993; Thorne 1995).  Proposed
``dual--recycled'' interferometers and spherical ``TIGA'' type resonant
detectors will be more sensitive than LIGO to the
higher frequency radiation, $\stackrel{{\textstyle >}}{\sim} 10^3$ Hz,
emitted during the brief final merger stage of the coalescence
(Meers 1988; Strain \& Meers 1991; Cutler et al.\ 1993;
Merkowitz \& Johnson 1994; Thorne 1995;
however, see Wilson \& Mathews 1995 who indicate that the
gravitational wave radiation emitted
during this phase may have a lower frequency than previously expected).
The gravitational wave radiation emitted during the merger phase in
double white dwarf binary evolution
is unlikely to be detected in the near future because the expected
frequency of the radiation falls in or
just beyond the upper end of the frequency range ($10^{-4}$--$10^{-1}$ Hz)
of proposed space--based
laser interferometric detectors (Faller et al.\ 1989; Hough et al.\ 1995)
and the duration of the phase will be too short to produce a significant
signal in this range of the detectors' sensitivity.

Because the final stages of binary coalescence are driven in part by
sizeable nonlinear tidal effects, numerical hydrodynamic techniques
must be used to properly follow the evolution of merging binaries.
The first step in performing such a hydrodynamic simulation
is the construction of an appropriate initial model.  The coalescence of 
the chosen binary system must proceed on a dynamical timescale
(on the order of a few initial orbital periods),
in order for an explicit hydrodynamics code to be able to carry out the simulation in
a reasonable amount of computational time.  Hence, the components of the initial
binary model must either be at a separation where they are dynamically 
unstable to coalescence or they must be forcibly
brought to coalescence 
from a wide separation (e.g.{}, by draining orbital 
angular momentum away from the system in a way that mimics the
effects of the gravitational wave radiation reaction).
Using the former methodology, the work presented herein focuses
on the identification of dynamically unstable
binary systems.

The initial separation at which a particular binary model becomes
dynamically unstable to merger, if one exists, can be found
via a stability analysis of a set of binary models
constructed in hydrostatic equilibrium, along a constant mass
sequence of decreasing orbital separation.  This sequence serves as 
an approximate representation of the evolution of the binary as
its components are brought closer together by the effects of gravitational
radiation.
Such analyses have recently been done by Lai, Rasio, \& Shapiro
(hereafter LRS) and by Rasio \& Shapiro (hereafter RS) for
binaries with polytropic equations of state (EOS).  In a polytropic
equation of state, the pressure $P$ is expressed in terms of the density $\rho$
as $P=K \rho^{1+1/n}$, where $K$ is the polytropic constant and
$n$ is the polytropic index (see \S 2).  
The analytical work of LRS utilized an 
approximate energy variational method and studied detached binaries with
components having various mass ratios, spins, and polytropic indices
(LRS 1993a, b, 1994a, b).
The numerical work of RS utilized the smoothed particle hydrodynamics
technique to study detached
and contact binaries with components having various mass ratios, but
equal spins and polytropic indices
(RS 1992, 1994, 1995; for earlier work see Gingold \& Monaghan
1979; Hachisu \& Eriguchi 1984a, b; Hachisu 1986b).

We
performed stability analyses of equilibrium sequences of
double white dwarf binaries constructed with the zero--temperature
white dwarf equation of state (Chandrasekhar 1967), double neutron
star binaries constructed with realistic neutron star equations of
state (adapted from Cook, Shapiro, \& Teukolsky 1994), and, for
the sake of comparison with the work of LRS and RS, polytropic binaries
with $n=0.5$, 1.0, and 1.5 equations of state.  The examined
equilibrium sequences were
constructed with the Self--Consistent Field technique of Hachisu (1986a, b),
which produces models of rotating, self--gravitating fluid systems
in hydrostatic equilibrium.  For simplicity, all binary models along these
sequences were constructed as synchronously rotating
systems
having equal--mass ($q=1.0$) components.
The relative stability of individual binary systems
along selected sequences was examined using a 
three--dimensional (3D), finite--difference hydrodynamics code.
Both the construction of our equilibrium binary sequences and our stability
tests along these sequences have been done using purely Newtonian gravity
and Newtonian dynamics.

Our numerical techniques are briefly described in
Section 2.  Constructed equilibrium sequences are presented in Section 3
and our dynamical tests of the stability of individual models along
selected sequences
are presented in Section 4.  Finally, the implications of these results are
discussed in Section 5.

\section{Numerical Techniques}
Our simulations of close binary systems involve the solution of the
following set of equations which govern the structure
and evolution of a nonrelativistic fluid in cylindrical coordinates:
 
\noindent
the continuity equation,
\begin{equation}
\frac{\partial \rho}{\partial t}+\nabla \cdot \left(\rho \vec{v}\right)=0;
\end{equation}
%
the three components of the equation of motion,
\begin{eqnarray}
\frac{\partial S}{\partial t}+\nabla \cdot \left(S \vec{v}\right) &=&
 -\frac{\partial P}{\partial R}-\rho \frac{\partial \Phi}{\partial R}+\frac{A^2}{\rho R^3}, \\
\frac{\partial
{\mathcal{T}}}{\partial t}+\nabla \cdot \left({\mathcal{T}} \vec{v}\right)&=&
 -\frac{\partial P}{\partial z}-\rho \frac{\partial \Phi}{\partial z}, \\
\frac{\partial A}{\partial t}+\nabla \cdot \left(A \vec{v}\right)&=&
 -\frac{\partial P}{\partial \phi}-\rho \frac{\partial \Phi}{\partial \phi};
\end{eqnarray}
%
Poisson's equation
\begin{equation}
\nabla^2 \Phi = 4 \pi G \rho;
\end{equation}
and the equation of state (see below).
In the above equations, $\vec{v}$ is the velocity;
$S=\rho u$, ${\mathcal{T}}=\rho w$, and
$A=\rho R v_{\phi}$ are the radial, vertical,
and angular momentum densities,
respectively (where $u$, $w$, and $v_{\phi}$ are the radial, vertical,
and azimuthal components of the velocity, respectively);
$R$, $\phi$, and $z$ are the cylindrical coordinates;
and $\Phi$ is the gravitational potential.

We have used three types of barotropic equations of state
in this work.  The first, and simplest,
type is a polytropic equation of state for which
\begin{equation}
P=K \rho^{1+1/n},
\end{equation}
where $K$ is the polytropic constant and $n$, the polytropic index, determines
the degree of compressibility of the fluid (the higher the value of $n$,
the more compressible/softer the fluid).

The second type of equation of state used is the zero--temperature 
white dwarf (WD) equation of state (Chandrasekhar 1967),
which represents the pressure distribution of a completely degenerate 
electron gas:
\begin{eqnarray}
P & = & a_{0}\left[x\left( 2x^2-3\right) \left(x^2+1\right)^{1/2} + 3\ln
\left( x+\sqrt{1+x^2}\right) \right] \nonumber \\
x & \equiv & \left( \rho /b_{0}\right)^{1/3},
\end{eqnarray}
where $a_{0}=6.00 \times 10^{22} \rm{dynes\:cm}^{-2}$, $b_{0}=1.95 (\mu_{e}/2)
\times 10^{6} \rm{g\:cm}^{-3}$, and $\mu_e$ is the mean molecular weight
per electron.  We have used $\mu_e = 2$ in all of our computations.  The heaviest
nonrotating single object that can be constructed with this equation
of state has a mass of
$1.44 M_{\odot}$ (this is the Chandrasekhar mass $M_{CH}$).

The third type of equation of state used here is a realistic neutron star (NS)
equation of state.  We have chosen 
three such equations of state 
(from among the fourteen realistic NS equations of state listed in Cook, Shapiro,
\& Teukolsky [1994, hereafter CST]),
each with a different degree of compressibility (one soft, one medium, and
one hard).
Specifically, the chosen soft equation of state is CST's 
equation of state F; the medium one is CST's equation of state
FPS; and the hard one is CST's equation of state L 
(see references
within CST for the original sources of these equations of state).
We obtained these equations of state 
in tabular
form from Cook (1995). The tables each provide 
$\sim500$ values of the pressure $P$ for values of 
$\rho$ ranging over 15 orders of magnitude, from \mbox{$\sim8\:{\rm g\:cm^{-3}}$} to 
$10^{16} \rm{g\:cm^{-3}}$ (note that it is actually the number density
$N=\rho/m_{neutron}$, where $m_{neutron}=1.67\times 10^{-24}\:{\rm g}$, that is tabulated).
Because we wanted to perform {\it{parallel}} 
finite--difference hydrodynamics (FDH) simulations of systems with these 
equations of state and
did not possess an interpolation algorithm designed for efficient use on a
parallel machine,
polynomial fits to the tabular data were necessary.  
Some numerical manipulation of the data was also needed because of the
particulars of the technique used in the initial model construction.
(See New [1996] for details.)

If the only motion of a fluid system is rotation about an axis
with an angular velocity $\Omega$, which is constant in time and a
function only of the distance from the rotation axis, 
the structure of the system is described by
the following single expression:
\begin{equation}
\frac{1}{\rho}\nabla P+\nabla \Phi + \nabla \Psi(R)=0,
\end{equation}
where the z--axis has been chosen as the axis of rotation and the 
centrifugal potential $\Psi(R)=-\int\Omega^2(R)R\,dR$.
Such a fluid is said to be in hydrostatic
equilibrium because the forces due to its pressure and to its
gravitational and centrifugal potentials are in balance.  All of the
initial equilibrium binary systems studied in this work
have been constructed in hydrostatic equilibrium
according to this prescription,
along with the additional constraint that angular velocity is a
spatial constant $\Omega_0$
(i.e.{}, not a function of $R$).  In this case of uniform rotation,
$\Psi(R)=-\Omega_{0}^{2} R^{2}/2$.

\subsection{Self--Consistent Field Code}
The method we have used to construct 
the equilibrium models is Hachisu's grid--based 3D,
Self--Consistent Field (HSCF) technique (Hachisu 1986a, b).  
This iterative
technique produces rotating, self--gravitating fluid systems in hydrostatic 
equilibrium.
Our version of the HSCF 3D code computes 
the gravitational potential via a direct 
numerical solution of Poisson's equation (5).
Details of the method used can be found in Tohline (1978).

An estimate of the quality of the converged equilibrium configuration
is obtained from a determination of how well the energy is balanced in
the system.  This balance is measured by the {\it virial error} $VE$:
\begin{equation}
VE\equiv \frac{\mid 2T+W+3\int P\,dV \mid}{\mid W \mid},
\end{equation}
where $T$ is the kinetic energy; $W$ is the gravitational potential
energy; and $V$ is the volume of the model.
The virial errors in our equilibrium models constructed with
polytropic and WD EOS were typically $\sim 10^{-3}$--$10^{-4}$;
those in models constructed with the realistic NS EOS were
typically $\sim 10^{-2}$.

The forms of the WD and realistic NS equations of state are such that
when they are used in the HSCF code, the density maxima
$\rho_{max}$ of the models to be constructed must be given to the code
as input.
Thus because we were interested in constant mass sequences
for our stability analyses of close binaries, we had to
perform an iteration in the choice of $\rho_{max}$ until we
arrived at a configuration with the desired $M_T$, in the case
of models with the WD and realistic NS equations of state.
However, in the polytropic case, converged
models can actually be obtained without an {\it a priori}
choice of $\rho_{max}$ and then later scaled as desired.

Our 3D equilibrium configurations are assumed to be symmetric about
the $z=0$ (equatorial) plane; this symmetry will be referred to as
equatorial symmetry.  Because our version of the 3D HSCF code constructs
binaries with only equal--mass components, a periodic symmetry over the
azimuthal range $0<\phi<\pi$ is also assumed.  This means that a quantity
$U$ specified at an angle $\phi$ is equivalent to that same quantity
specified at all angles $\phi^{\prime}$ for which $\phi^{\prime}=
(\phi+m\pi)$ and $m$ is an integer:
\begin{equation}
U(\phi+m\pi)=U(\phi).
\end{equation} 
This symmetry will be referred to as $\pi$--symmetry.

\subsection{Finite--Difference Hydrodynamics Code}
A finite--difference hydrodynamics (FDH) code was used to solve,
on a discrete numerical grid, equations (1-5) and an equation of state
(see above),
which govern the temporal evolution of a fluid.
FDH codes differ from smoothed particle hydrodynamics (SPH)
codes in that they follow the
evolution of the fluid as it flows through a fixed set of grid cells,
instead
of treating the fluid as a set of particles and following the
evolution of each particle.

The 3D FDH code used in this present study is a Fortran 90 version
of the 
Fortran 77 code described by Woodward (1992) and Woodward, Tohline,
\& Hachisu
(1994). It was written, principally by Woodward, to take 
advantage of the parallel
architecture of the MasPar computers on which it is run.  The accuracy of the
code is second order in both time and
space.  The numerical techniques
employed are discussed in detail in Woodward (1992).
The solution to Poisson's equation (5) is obtained through the
ADI method (Cohl et al.\ 1997).

As in the HSCF code,
the grid cells are uniformly spaced in each of
the three directions and equatorial and $\pi$--symmetries are assumed.
The single precision hydrodynamic simulations presented here
were performed
on cylindrical grids with resolutions of $64\times 64\times 64$.

In the binary stability analysis simulations presented in \S 4,
from $\sim 2,600$ to 16,700 timesteps were required to follow
each binary
through one initial orbital period $P_I$.
On the 8K--node MasPar MP--1 at LSU, each timestep took $\approx 19$ CPU
seconds (or $\approx 73\:\mu{\rm sec}$ per grid zone); hence,
these simulations required $\approx 14$ to 88 CPU hours per $P_I$.
This large range is
due to the variation in the size of the integration timestep that
could be taken
in the different simulations.  The size of this timestep is
restricted in order
to ensure the numerical stability of the computations. 
The simulations we performed varied in length from
$1\:P_I$ to $5\:P_I$.
A few of our binary dynamical stability test
simulations were run on the 4K--node MasPar MP--2 at the 
Scalable Computing Laboratory of the DOE Ames Laboratory at
Iowa State University.
The CPU time per timestep required for simulations conducted on the MP--2
is about $3/5$ that
of the time required to run on the MP--1 at LSU. 

Our FDH code typically follows the fluid evolution in the inertial
reference
frame.  However, we chose to incorporate the option of running
the code in a
frame of reference which rotates with the initial angular velocity
of the
fluid, $\Omega_{0}$.  This choice was motivated by a desire to minimize 
numerical effects which might artificially influence the stability
of the
binary systems studied.  The particular effect we sought to minimize was
dissipation due to numerical viscosity, which arises from the coarseness
of the finite differencing.  The hope was that diminishing the motion of
the fluid through the grid by running in the rotating reference frame
would also diminish the dissipative effects of numerical viscosity
on the fluid (see \S4 for further details).
(We also tried updating the angular velocity of the rotating frame of 
reference once during some of the simulations presented here,
in order to further
minimize the dissipation due to numerical viscosity).

The rotating reference frame adds two terms to the radial
equation of motion (2) and one to the azimuthal equation (4) 
(cf.{}, Norman \& Wilson 1978): 
\begin{eqnarray}
\frac{\partial S}{\partial t}+\nabla \cdot \left(S \vec{v}\right) &=&
 -\frac{\partial P}{\partial R}-\rho \frac{\partial \Phi}{\partial R}+\frac{A^2}{\rho R^3}
 + \rho \Omega_{0}^2 R + \frac{2 \Omega_{0} A}{R}, \\
\frac{\partial A}{\partial t}+\nabla \cdot \left(A \vec{v}\right)&=&
 -\frac{\partial P}{\partial \phi}-\rho 
\frac{\partial \Phi}{\partial \phi}-2\Omega_{0} R S.
\end{eqnarray}
The $\rho \Omega_{0}^2 R$
term that has been added to the radial equation of motion
results from the centrifugal force; the other two added terms
result from
the Coriolis force.  Note that the centrifugal term in
equation (11) can be
rewritten as ${A^{\prime}}^2/(\rho R^3)$,
where $A^{\prime}\equiv \rho \Omega_{0} R^2$.
We use this form in the actual computation of the centrifugal term, 
with $A^{\prime}$
centered at the same place in each grid cell as is $A$, in order to 
be numerically
consistent with the computation of the curvature term
$A^2/(\rho R^3)$ in
the radial equation of motion.

A discussion of the boundary conditions, vacuum treatment, and
rotation axis treatment implemented in our hydrodynamics code
is presented in New (1996).

\section{Equilibrium Sequences}

We have constructed hydrostatic equilibrium sequences of synchronized
close binaries with polytropic as well as realistic white dwarf and neutron
star equations of state.  The individual binary models along each sequence
have the same equation of state and constant total mass $M_{T}$, but decreasing
binary separation $a$. Here $a$ is the distance measured between the
pressure [density] maxima of the stellar components.
Each such sequence represents a quasistatic approximation 
to the evolution of a binary system in which gravitational radiation 
gradually carries away the system's
orbital angular momentum.  These binary models were constructed, on
$128\times 128\times 128$ grids, with
the HSCF technique (see \S 2.1), which creates models of rotating,
self--gravitating fluid systems in hydrostatic equilibrium.

It should be noted that the true physical viscosity present in double neutron star
binaries is not expected to be strong enough to enforce synchronization
(Bildsten \& Cutler 1992; Kochanek 1992) and the viscosity of the
degenerate material in double white dwarf binaries probably is not
strong enough to synchronize them either (this can be shown by applying
the arguments given in Bildsten \& Cutler 1992 to white dwarfs and using
the values for the viscosity of degenerate material given in Durisen 1973).
However, if magnetic fields are present, they may bring about synchronization.
In any case, synchronization is at least a simplifying assumption.
Furthermore, since neutron stars have relatively strong gravitational
fields, Newtonian models and simulations of double neutron star binaries
need to be viewed with caution.

\subsection{Polytropic Sequences}
The equilibrium sequences that we constructed
for binary models with polytropic indices
$n=0.5$, 1.0, and 1.5 are displayed in Figure 1.  For each $n$,
the total energy $E$ and total angular momentum $J$ are plotted
versus the separation $a$.
Note that we do not claim that the results given in this figure or in the
rest of the manuscript are necessarily accurate to the number of digits
in which they are reported; the number of digits in which the results are
presented allows the display of characteristics of the equilibrium
sequences and the differentiation between individual models on these sequences.
As mentioned in \S 2.1, the equations of state 
of the polytropic models are such that the total mass of the system $M_T$
does not have
to be chosen before its construction, but can be scaled 
afterwards as desired.

There are in fact three parameters, $M=1/2M_T$, $R_{Sph}$, and the
polytropic constant $K$, which set the scale of a polytropic model.
Here $R_{Sph}$ is the radius of a spherical star of mass $M$ and
polytropic index $n$.  These parameters are related according to the following 
equation (Chandrasekhar 1967):
\begin{equation} 
M=4\pi m_{n}\left[\frac{(n+1)K}{4\pi G}\right]^{\frac{-(3+n)}{2(1-n)}}
\left[\frac{R_{Sph}}{r_n}\right]^{\frac{(3-n)}{(1-n)}},
\end{equation}
where $m_n$ and $r_n$ are Lane--Emden constants for a particular value of $n$
(see Table 1 for their values corresponding to $n=0.5$, 1.0, and 1.5).
Thus, only two of these three parameters are independent; if
two of them are specified, the other one is automatically determined.
The quantities
in Figure 1 are themselves normalized to $G$, $M$, and $R_{Sph}$.
Specifically, $E$ has been divided by $(GM^{2}R_{Sph}^{-1})$;
$J$ has been divided by $(G^{1/2}M^{3/2}R_{Sph}^{1/2})$; and $a$
has been divided by $R_{Sph}$.

In the discussion that follows, as in Figure 1,
all values of the binary separation $a$ 
have been normalized to $R_{Sph}$.
On each sequence, Point C marks the system where the surfaces of the 
two binary components just come into contact.  Systems to the right of
this point are detached binaries and systems to left are contact binaries
or ``dumbbells.''  
For the sake of illustration, Figure 2 displays an isodensity surface image
of an example detached binary model ($a=3.28$, $n=1.0$)
and of an example dumbbell model ($a=2.70$, $n=1.0$).
Points M mark the models along each sequence which have
the minimum total energy and
the minimum total angular momentum.  Along all 
three polytropic sequences,
the minimum in $E$ occurs at the same separation as the minimum in $J$.
(See \S 3.4 below for a discussion of the significance of these minima.)
The model with the smallest separation on each sequence, marked with a T, will
be referred to as the
``terminal'' model.  No synchronously rotating
binary models with the equation of state
particular to that sequence
can be constructed in equilibrium with a smaller separation
than that of the terminal model because
the centrifugal force would exceed the gravitational force along the
equator of such systems.

The separation at which this termination of the sequence occurs increases
from $a=2.45$ for $n=1.5$ to $a=2.76$ for $n=0.5$.  The separation at which
the minima occur also increases from $a=2.70$ for $n=1.5$, to $a=2.89$
for $n=1.0$, to $a=3.11$ for $n=0.5$.  Note that contact occurs to the right
of the minima for $n=1.5$ and to the left of the minima for $n=1.0$ and 0.5.  We
have determined that Points C and M coincide for $n=1.177$.
If the binary components were spherical, their separation at the point of
contact would be $a=2$.  However, $a({\rm C})$ ranges from 2.81 for $n=1.5$ to
2.94 for $n=0.5$.

\subsubsection{Comparison with Previous Work}

In this section, we compare our polytropic equilibrium sequences
to those of LRS (1993b) and RS (1992, 1995).  Note that LRS (1993b)
and RS (1992, 1994, 1995) define binary separation as the distance between the
centers of mass of the binary components and, as mentioned above, we
define it as the distance between the pressure (density) maxima of
the components.  For ease of comparison, the values of binary separation
presented
in this section that are based on our work do represent the separation
between the centers of mass of our binary components.
In this section and the rest of this manuscript, any binary separation which
refers to the distance between the components centers of mass will be denoted
as $a_{cm}$.

The analytical stability analyses of LRS (1993b), which utilized an
approximate energy variational technique, also showed that simultaneous
minima in $E$ and $J$ existed along sequences of constant mass,
synchronized binaries with $n=0.5$, 1.0 and 1.5.  (According to LRS 1993b,
the minima occur at
$a_{cm}=2.99$ on the $n=0.5$ sequence and at $a_{cm}=2.76$ on
the $n=1.0$ sequence.)
This analytical method cannot construct contact binaries and thus, according
to our sequences, should not be able to identify minima on the $n=1.5$
sequence.  However, because the $n$ at which the minima and points of contact
coincide is $\sim 2.0$ in their study, LRS (1993b) do find minima for the $n=1.5$ sequence
at $a_{cm}=2.55$.

In addition, approximate equilibrium sequences were constructed with the
SPH techniques of RS (1992, 1995).  The $n=1.0$ sequence presented in
LRS (1993b) has simultaneous minima at $a_{cm}=2.90$, which is closer to
our value of $a_{cm}=2.98$ than the analytically determined $a_{cm}=2.76$ of
LRS (1993b).  The $n=1.5$ sequence presented in RS (1995) has the minima
at $a_{cm}=2.67$.  No sequence with $n=0.5$ has been published by RS.
Table 2 contains a summary of the separations $a_{cm}$ of the models at the 
minima of the polytropic
sequences as determined in LRS (1993b), RS (1995), and this work.
For completeness, this table also gives the values of binary separation
determined
by this work in terms of the separation $a$ between the pressure maxima
of the components.

Hachisu (1986b) also shows equilibrium sequences of $n=0.5$ and 1.5 polytropes.
However he presents his results as sequences of $\Omega_0^2$ versus $J^2$ instead
of $E$ or $J$ versus $a$.  A comparison between Hachisu's (1986b) 
results and ours is given in Figure 3.
To conform with Hachisu's notation,
the quantities in this figure
are normalized to $4\pi G$, $M_T$, and $V_T$, where $V_T$ is the total
volume of the binary system.
Specifically, $\Omega_{0}^{2}$ has been divided by $(4\pi GM_{T}/V_{T})$
and $J^2$ has been divided by $(4\pi GM_{T}^{3}V_{T}^{1/3})$.

\subsection{White Dwarf Sequences}
We have constructed equilibrium sequences for binary models with the zero--temperature
WD equation of state.
Because the HSCF technique requires that the maximum density of the
desired model (which sets $M_T$) be given as input when this equation
of state is used,
it is impossible to build a single sequence that can be scaled to any desired
$M_T$ with this equation
of state.  Instead we have constructed separate WD sequences, each
of which represents models with one specific value of $M_T$.  We have constructed nine such
sequences with $M_T$ ranging from .298 $M_\odot$ to 2.72 $M_\odot$.  Four
representative sequences, with $M_T=.500$, 1.19, 2.03, and 2.72 $M_\odot$, are
shown in Figure 4.  The other five sequences are displayed in Appendix A,
along with
the four WD sequences presented in Figure 4.
(In addition, two WD sequences with $M_T\:[$near $2M_{CH}] = 2.81$ and 
2.85 are presented in Appendix A; 
they have been excluded from the discussion below because of their irregular nature.)
The normalization in Figure 4 is the same as that in Figure 1.  However, in this
case $R_{Sph}$ has been determined numerically by constructing a spherical
white dwarf of mass $M=M_{T}/2$ in hydrostatic equilibrium.

The separations $a$ of the models at the points of contact, the minima,
and the terminal points on the constructed white dwarf equilibrium
sequences are shown in Figure 5 as a function of the total binary system mass.
Along the WD sequences, the separation of the terminal model
gradually increases
from $a=2.45$ when $M_T=.298\:M_{\odot}$ to $a=2.86$ when $M_T=2.72\:M_{\odot}$.
As in the polytropic sequences presented in the previous section,  simultaneous
minima in $E$ and $J$ exist along each WD sequence.  The point of contact
on these sequences always occurs at a larger separation than the minima; the
separation at which it occurs also gradually increases from $a=2.81$ for
$M_T=.298\:M_{\odot}$ to $a=3.05$ for $M_T=2.72\:M_{\odot}$ (except for a slight
decrease in this separation for the $M_T=1.19\:M_{\odot}$ sequence).
The separation of the model at the minimum of each sequence also increases
from $a=2.70$ for $M_T=.298\:M_{\odot}$ to $a=2.97$ for $M_T=2.72\:M_{\odot}$;
however, in this case there is a somewhat more substantial decrease in this
separation between the $M_T=.500\:M_{\odot}$ and $M_T=2.36\:M_{\odot}$ sequences
(for which $a=2.73$).

\subsubsection{Comparison with Previous Work}

Hachisu (1986b) has constructed double white dwarf
binary sequences along which $\rho_{max}$,
instead of $M_T$, was held constant.  Figure 6 shows a comparison between
the $J$ versus $M_T$ relations for the models with the minimum angular momentum
on the Hachisu (1986b) sequences and those on our sequences.  The 
angular momentum in this figure is normalized to $10^{50}$ in cgs units and the mass is
normalized to $M_\odot$.  The comparison is excellent for low masses, however
the relations deviate slightly for $M_{T} \geq 2\:M_{\odot}$.

\subsection{Neutron Star Sequences}

We have constructed equilibrium sequences for binary models with three
realistic NS equations
of state of varying compressibility (F, soft; FPS, medium; L, hard).
As in the case of the WD equation
of state, the desired $M_T$ for each of these sequences
must be specified prior to their construction.  We have chosen to construct
one sequence with $M_T=2.80\:M_{\odot}$ for each of the three equations of state.
These three sequences
are displayed in Figure 7 and have been normalized to $G$, $M$, and $R_{Sph}$. 
The values of $R_{Sph}$ were determined numerically by constructing a
spherical star of mass $M=M_{T}/2$, in hydrostatic equilibrium, with
each of the three NS equations of state.

The terminal model occurs at the same separation on the F and
FPS sequences ($a=2.59$) and at a slightly
wider separation ($a=2.62$) on the L sequence.  The locations of the
minima in $E$ are not very well defined on these sequences.
For the F sequence, the $E$ minimum likely occurs in the range
$2.75<a<2.81$; for the FPS sequence, in the range $2.72<a<2.81$;
and for the L sequence, in the range $2.72<a<2.84$.
The $J$ minima are well defined and occur at nearly the same separation
on all three sequences (on the F sequence at $a=2.80$, on the FPS sequence at
$a=2.78$, and on the L sequence at $a=2.79$).
Like the white dwarf sequences, the minima always occur at a smaller
separation than the points of contact.  However, on all of the neutron star
sequences these two points are very close together.  
The separation between points C and M thus seems to be determined by the 
lower density regions of the objects since this is where the equations
of state, which differ 
significantly only
in the density regimes above nuclear density
($2.8\times 10^{14}\:{\rm g\:cm}^{-3}$), are similar. 

The scatter in $E$ displayed near the minima of these sequences
may result in part from our piecemeal reconstruction of
Cook's (1995) tabular NS equations of state (see \S 2)
and from the approximate
manner in which we calculate the internal energy $E_{int}$
of the models (we compute an effective polytropic index $n_{eff}$
for each grid zone in the model and then use these spatially varying
indices to calculate the internal energy according to
$E_{int}=\int n_{eff} P\,dV$).
Because of these
factors, we identify the $J$ minima as the true minima along these sequences
and, as before, have marked their location with the letter M.
Another factor which must be kept in mind when studying the features of these
NS sequences is that the virial error ($VE$), which measures the quality of the
equilibria, is $\sim 10^{-2}$.
This is one to two orders of magnitude higher than the $VE$ of models along
the polytropic and WD equation of state
sequences and almost certainly results in part from the
piecemeal forms of the NS equations of state used.
Note that for the polytropic and WD equation of state,
we typically very accurately pinpointed the
location of the miminum by moving point B (one of the two points on the
surface of the star that must be given as input to the HSCF code
and whose position influences the separation of the binary components)
one grid cell at a time in the
region of the minimum in order to get as many models as possible with the
grid resolution we used.  However, since we had difficulty obtaining
converged models for some portions of the NS sequences, we moved point B
two grid cells at a time when constructing NS models.
Thus the locations of the minima are slightly
more uncertain for the NS equations of state than in the other studied
equations of state.  

\subsection{Nature of the Minima}

Turning points on equilibrium sequences, like the
minima present in $E$ and $J$ along the sequences presented
above, are usually associated with
points of instability.  The two types of instability
which will be discussed below are {\it secular} instability
and {\it dynamical} instability.
For an instability to be classified
as dynamical, according to the convention of LRS (1993b),
it must conserve energy, angular momentum, and
circulation. If an instability proceeds as a result
of some mechanism that dissipates one of the conserved quantities, 
e.g.{}, via viscosity or gravitational
radiation, then it is classified as a secular instability
(see also Tassoul [1978] for discussions of dynamical versus secular
instabilities).
A dynamical instability takes place on the dynamical timescale
of the system; a secular instability takes place on the
timescale of the particular dissipative mechanism which induces it.

The minima in $E$ and $J$ on synchronous binary sequences
have been identified as points of secular instability by previous authors. 
LRS (1993b) stated that the neighboring models adjacent to the 
model at the minimum of each sequence are uniformly rotating and
therefore can only be reached with the aid of viscosity.
Thus they concluded that this minimum cannot be associated with a
dynamical instability since viscosity does not preserve circulation.
Their approximate analytical method predicts that 
on polytropic sequences with $n<0.7$, dynamical instabilities
exist at separations smaller than those of 
the minima.  For the $n=0.5$ sequence, these authors conclude that the 
dynamical instability sets in at $a_{cm}=2.68$.
Hachisu (1986b and references therein) also labels turning points
along his synchronous polytropic and WD sequences as points of secular
instability.

Because our hydrodynamics code does not include the dissipative
effects of either the gravitational radiation reaction or the true
fluid viscosity, we are unable
to confirm the presence of a secular
instability with a hydrodynamics simulation.  However,
it is possible to study the dynamical stability of a system
with our code.
In the following section, we present the
results of our FDH tests of the dynamical stability of models
on sequences selected from those presented above and compare these
results with those of other authors.

\section{Hydrodynamic Tests of Stability}
To determine if a point of dynamical instability exists on an equilibrium
sequence like those discussed in the previous section, the dynamical stability
of individual models along the sequences may be tested with FDH simulations.
We have done just that for the three polytropic sequences of \S 3.1;
for a low mass ($M_T=.500\:M_{\odot}$) and a high mass ($M_T=2.03\:M_{\odot}$)
WD sequence from \S 3.2; and for two of the three realistic NS equations
of state sequences of
\S 3.3.

All of these stability tests were performed in the rotating reference frame
(\S 2.2) in order to minimize the dissipative effects of the numerical
viscosity which results from the discrete nature of the computational simulation.
The influence of numerical viscosity on the evolution of binary systems
can be seen in Figure 8 which shows a comparison between simulations
of one particular WD binary system ($M_T = .500\:M_{\odot}$ and
$a=2.63$) performed in the inertial frame (dashed curve) and in the rotating
frame (solid curve).  This figure shows the evolution of the moment of
inertia of the system $I$, as a function
of time $t$.
The evolution of $I$ is more instructive than the evolution
of the binary separation itself since the latter quantity can only be
measured by discrete jumps in the location of the density maximum on
the grid, whereas $I$ varies smoothly with time.

As can be seen from Figure 8, the binary appears to be dynamically
unstable when the simulation is performed
in the inertial frame (dashed curve), 
as $I$ plummets on a timescale of $2$--$3\:P_I$.
By contrast, the same model evolved in the rotating frame is certainly not
unstable to merger, as can be seen by the relatively steady behavior of its
moment of inertia over time.  Thus, simulations done in the rotating frame
prevent the
misidentification of models as dynamically unstable, which are so only
because of numerical artifacts.  In order to illustrate that
the accuracy of the finite differencing scheme also influences the
amount of numerical viscosity present in a simulation, Figure 8
also shows the same model evolved in the rotating frame but with a FDH code that
used a first--order
accurate advection scheme to compute the divergence terms in equations (1-4)
(dot-dashed curve).  This figure indicates that the accuracy of
the code has an effect on the evolution of this system that is similar to
that of the flow of the fluid through the grid in the inertial frame.

Models along the equilibrium
sequences presented in \S 3 were constructed with a grid resolution
of $128\times 128\times 128$.  However, a FDH simulation with this 
resolution cannot be done on the MP--1 computer at LSU.  So portions of the
selected sequences mentioned above have been recomputed on
$64\times 64\times 64$ grids.  It is the stability of models on these
new sequences which actually has been tested. 
For completeness, Table 3 gives
the separations of the points of contact, minima, and
terminal models for both the $64^3$ and the $128^3$ versions of
the polytropic and WD sequences discussed below.
(We do not show this comparison for the NS sequences because
we have not determined or were unable to accurately determine
the location of some of these points on the $64^3$ NS sequences.)
\subsection{White Dwarf Sequences}
\subsubsection{${{M_T = 2.03\:M_{\odot}}}$}
In the lower panel of 
Figure 9, $I(t)$ is shown for WD binaries ($M_T = 2.03\:M_{\odot}$)
with separations ranging
from $a=2.80$ (triple dot--dashed curve,
a dumbbell model just past the point of contact)
to $a=2.53$ (solid curve; the terminal model on the sequence).
For the sake of convenience, the relevant
equilibrium sequence itself is reprinted in the
upper panel of Figure 9.
The moments of inertia of the binary models at the points of contact,
minima, and termination along the equilibrium sequence (constructed
on a $128\times128\times128$ grid) are labelled
with the letters M, C, and T, respectively.
Note that
because the binary models used in the FDH stability tests were
from sequences constructed on $64\times64\times64$ grids, there may be a slight
offset between the initial values of $I$ for these models and the points
marked as M, C, and T on the $I(t)$ plot since they correspond to the
$128\times128\times128$ sequence.

All of the models tested on the sequence
are dynamically stable.  Thus we conclude that no point of dynamical
instability exists along this sequence.
In Figure 10, we present isodensity images of an example
of the evolution of a stable binary.  The images are from the simulation
of the model at the minimum of the $M_T=2.03\:M_{\odot}$ sequence, 
corresponding to the dot--dashed curve of Figure 9.  Since the simulation
was performed in the rotating frame, the binary does not appear to pivot
very much in this set of images even though the simulation was
carried out for $\sim 4P_I$.
\subsubsection{${{M_T = .500\:M_{\odot}}}$}
Our tests of the $M_T = 2.03\:M_{\odot}$ sequence began with models
of wider separation and continued up the sequence to the terminal model.
Because no point of dynamical instability was found on this sequence,
we performed our tests of the $M_T = .500\:M_{\odot}$ sequence
on models much closer to the terminal model, knowing that we could work our
way back down the sequence to the point of dynamical instability if we
found models which were unstable.  (This methodology assumes that
all models which have separations less than that of the model
located at the point of dynamical instability will also be unstable.)
But, as can be seen from
the bottom panel of Figure 11, both a model at $a=2.63$ (dashed curve) and the
terminal model at $a=2.45$ (solid curve) are dynamically stable against merger.
Thus we again infer that there is no point of dynamical instability along 
this sequence.
\subsection{Polytropic Sequences}
\subsubsection{${{n=1.5}}$}
Based on the results of the stability tests of the WD sequences,
we began our investigation of this sequence with the terminal model
($a=2.45$).
Because an $n=1.5$ polytrope is supposed to be a fair representation
of a low mass WD, we have plotted the $I(t)$ of the terminal
model ($a=2.45$) in Figure 11 (dot--dashed curve) along
with the models of the low mass WD sequence.
The polytropic model has been scaled to represent a binary
with $M_{T}=.500\:M_{\odot}$ and
a $R_{Sph}$ equal to that of a spherical WD with the same mass.
This sets $K$ according to equation 13.  As can be
seen, the evolution of the $n=1.5$ terminal model closely follows
that of the low mass WD terminal model.  From its stability,
we infer that the remainder of the models on the $n=1.5$ sequence
are also stable and thus no point of dynamical instability exists
along this sequence either.  This result conflicts with the SPH
simulations of RS (1995) which identified a point of dynamical 
instability at $a_{cm}\simeq 2.4$ on their $n=1.5$ sequence.
\subsubsection{${{n=1.0}}$}

As Figure 12 illustrates, we have performed dynamical stability tests
of four binaries on the
$n=1.0$ sequence with separations ranging from $a=3.28$ (triple dot--dashed
curve)
to $a=2.62$ (solid curve; terminal model).  All were stable against merger on
a dynamical timescale.  The moment of inertia of the terminal model
actually increases by about 20\% over the timescale $t=4P_I$ depicted here;
the moment
of inertia of the initially most widely separated model ($a=3.28$)
decreases by about the same percentage.  The two other models exhibit
behavior in between these two extremes.  Note that after an initial dip,
the moment of inertia of the model at the minimum of the sequence
($a=2.88$, the dot--dashed curve) is nearly constant.
These results conflict with the SPH simulations of RS (1992) which
identified a model with $a_{cm}=2.8$ as being dynamically unstable.
\subsubsection{${{n=0.5}}$}

As Figure 13 illustrates, we have tested the dynamical stability of
five binaries on the $n=0.5$ sequence with separations ranging from
$a=3.41$ (long dashed curve) to $a=2.77$ (solid curve; terminal
model).  Both the model at the minimum of the sequence ($a=3.10$--short dashed
curve) and the terminal model were unstable to merger on a dynamical
timescale.  The other three more widely separated models,
including the model ($a=3.17$--dot--dashed curve) located just prior to
the minimum of the sequence, were stable.
Thus dynamical instability sets in at the minimum of this sequence
($a=3.10$, $a_{cm}=3.20$).  The SPH simulations of RS (1994) 
identified a point of dynamical instability along this sequence at
$a_{cm}=2.97$.

\subsection{Neutron Star Sequences} 
\subsubsection{Equation of State L}
As Figure 14 illustrates, we have tested the dynamical stability of
a NS binary system near the minimum
($a=2.79$) and a more widely separated model ($a=3.26$)
on the realistic neutron star equation of state L sequence.
The model near the minimum of the sequence became unstable to dynamical merger
in one $P_I$. The more widely separated model appears to be stable; its
behavior resembles that of the most widely separated models tested on
the $n=1.0$ and 0.5 sequences (see Figures 12 and 13).
Although we have not accurately identified the separation at which
the dynamical instability sets in along this sequence, it has obviously
already done so  at $a=2.79$.  By analogy with the $n=0.5$ sequence,
it is possible to infer that the onset 
of dynamical instability is associated with the region of the
minimum energy and angular momentum along the sequence.
\subsubsection{Equation of State F}
As Figure 15 illustrates, we have tested the dynamical stability of
the model at the minimum ($a=2.80$)
of the realistic neutron star equation of state F sequence.
Like the model near the minimum of the equation of state
L sequence, the model at the minimum
of this sequence is also unstable to dynamical merger.
Although we have not tested the stability of a widely separated model on
this sequence,
we infer, by analogy with the $n=0.5$ and equation of state L sequences,
that there exist more widely separated models on this sequence
that are stable and that the point of onset of the dynamical instability
may be associated with the region of the minimum.
\subsubsection{Equation of State FPS}
We have not tested the stability of any of the models on the realistic
NS equation of state FPS sequence.  However, since our tests of the stiff
equation of state L
and the soft equation of state F produced similar results,
we expect that the stability properties of the
models on this equation of state of medium stiffness will resemble
the properties of models on
these other realistic NS equations of state sequences.

\section{Discussion and Conclusions}
We have examined the dynamical stability of synchronously rotating,
equal--mass binaries with polytropic, zero--temperature white dwarf,
and realistic neutron star equations of state.
Specifically, we tested the dynamical stability of individual
models constructed along equilibrium sequences of binaries with
the same total mass $M_T$ and equation
of state but decreasing separation, in order to determine
if any models on these sequences were unstable to merger on a dynamical
timescale.

Our stability analyses started with two white dwarf (WD) equation
of state sequences, one with a
low total mass ($M_T=.500\:M_{\odot}$) and one with a fairly high total mass
($M_T=2.03\:M_{\odot}$),
used as representatives of the nine regular WD equation of
state sequences we constructed
(which ranged in mass from $M_T=.298$ to $2.72\:M_{\odot}$).  Our simulations
indicate that no points of dynamical instability exist on either of these
two sequences.  We have inferred from this result, that it is likely
(although not certain) that the other WD sequences are also dynamically
stable.  This being the case, we expect that
WD mergers will only happen via secular processes and that
it will not be possible
to properly simulate the coalescence of
equal--mass double white dwarf
binaries using explicit FDH (or SPH) techniques.  It
still remains to be seen whether or not
double white dwarf
binaries having unequal mass components
are susceptible to merger on a dynamical timescale.

Our examination of the $n=1.5$ and $n=1.0$ polytropic sequences
also identified no points of dynamical instability.  This result
conflicts with the published results of RS (1992, 1995) who identified a
dynamically unstable binary between the minimum and
the terminal point of the $n=1.5$ sequence and another just past the minimum of
the $n=1.0$ sequence.  However, like RS (1994), our
test of the $n=0.5$ sequence does indicate the presence of a dynamical
instability.  In our simulations, this instability sets in at the
minimum of the sequence.
RS (1994) locate the point of dynamical instability at $a_{cm}=2.97$.
Although they do not state where the minimum of their sequence is located,
the analytical work of LRS (1993b) places it at $a_{cm}=2.99$.  Recall that
LRS (1993b) label this minimum as a secular instability and they predict that
a dynamical instability sets in at a smaller
separation ($a_{cm}=2.68$) on this sequence.
However, in light of the simulations of RS (1994) and of this work, it 
seems likely that the minimum itself may be associated with the
onset of the dynamical instability.

The differences between our results and those of RS (1992, 1995) for
the sequences with the softer equations of state may result from the fact that SPH
has difficulty modeling low density regions that are more extensive
in systems with softer equations of state.  It is interesting to note that the
analytical work of LRS (1993b) predicts that points of dynamical
instability exist only on sequences for which $n$ is less than
$\approx 0.7$.
It would be intriguing to test
the stability of other polytropic sequences with our FDH code in order
to determine at what value of $n$ the dynamical instability first appears.
Another point that is worth emphasizing again, is that our stability
tests were run in the rotating frame of reference.  We believe that
we would have misidentified some truly stable models as being unstable had
these tests been performed in the inertial frame (see the discussion in
\S 4 associated with Figure 8).

Our stability analyses of the realistic neutron star (NS) equations
of state sequences identified
models at (or near) the minima of the soft equation of state F and
the hard equation of state L as being dynamically unstable.  However, because of
computing resource
constraints we have been unable to determine whether or not the
{\it onset} of the dynamical instability takes place in the regions of the
minima of these sequences.  If the $n=0.5$ sequence is taken as an
example, one might infer this to be the case.  Although we did not
test the stability of the medium equation of state FPS sequence we constructed,
we expect that such tests would yield results similiar to those of
the other two NS equations of state.
If further simulations of the $n=0.5$ and NS equations of state sequences do identify
the minimum as the point of dynamical instability, the question will arise
as to why this turning point does not also mark the onset of dynamical
instability on the sequences with softer equations of state.  At this time we are
unable to provide a physical explanation for this possibility
(see, however, the discussion in LRS [1993b]).

All of the binary models
that we have identified as stable against dynamical merger in \S 4
have not necessarily become steady--state configurations by the end of
our simulations.
In fact, at the end of our simulations the moments of inertia of some
of the stable models are still decreasing gradually and those of others
are still increasing gradually.
However, these binaries did not
become unstable to merger on a dynamical timescale (i.e.{}, $\sim$ a few
$P_I$).

It is interesting to note that this gradual (secular--type)
evolution of our models
is such that models before the minimum of a sequence (i.e.{}, models
with separations larger than that of the model at the minimum) tended
to exhibit decreasing moments of inertia and those after the minimum tended
to exhibit increasing moments of inertia. (This trend appears in the simulations
of models on the low mass white dwarf sequence and is more pronounced
in the tests of models with stiffer equations of state.)
LRS (1994b) have indicated that viscosity does indeed cause the orbits
of models before the minimum of a sequence to decay.  They also suggest
that as a result of the action of viscosity, the orbit of a binary past
the minimum ``\ldots can either expand \ldots as the system is driven to
a lower energy, stable synchronized state, or it can decay \ldots as
the stars are driven to coalescence.''

The LRS (1994b) description of the action of viscosity is consistent
with our results if the gradual evolution of the moments of inertia
of our models is attributed to the effects of the remaining numerical
viscosity in our code (which acted in our case to {\it {increase}} the moments
of inertia of dynamically stable models past the minima).  By carrying
out simulations in the rotating reference frame
(as is illustrated in Figure 8),
we have overcome the primary effect of numerical
viscosity, which is to dissipate orbital motions and drive systems to
coalescence.  In simulations performed in the rotating frame, it may be
a higher order effect of numerical viscosity
on any internal motions that develop in the systems that causes
the observed secular--type
evolution of our binary models.
Note that this higher order effect of numerical viscosity cannot be
identified as the mechanism responsible for driving any of our unstable
models to coalescence because
the timescale of their evolution was dynamical
and not gradual.  In addition, had this aspect of numerical viscosity
played a significant role in the evolution
of the terminal model on the $n=0.5$ sequence, its moment of inertia
should have increased gradually, but instead its moment of inertia
decreased and the model merged on a dynamical timescale.

Even though our work indicates that there is no point of dynamical
instability along the $n=1.0$ sequences, it would still be possible 
for an explicit hydrodynamics code to
follow the merger of a close binary with this equation of state if it was assumed
to represent a neutron star--neutron star
system.  This is because the timescale,
$\tau$, on which gravitational radiation
would drive a close neutron star--neutron star binary to coalescence
is on the order of its initial orbital period, $P_I$.
According to Shapiro \& Teukolsky (1983),
a point mass approximation to this timescale is
\begin{equation}
\tau =\frac{5}{256} \frac{c^5}{G^3} \frac{a^4}{2M^3}.
\end{equation}
If the binary at the minimum of our $n=1.0$ sequence is assumed to have
components with $M=1.4\:M_{\odot}$ and $R_{Sph}=10\:{\rm km}$,
then $\tau =2.5\:{\rm ms}=1.6\:P_I$.  (Note that for the binary at
the minimum of our $M_T=2.03\:M_{\odot}$ WD equation of state sequence,
$\tau P_{I}^{-1}=2.2\times 10^7$ and for the binary with $a=2.63$ 
on our $M_T=.500\:M_{\odot}$ WD equation of state sequence,
$\tau P_{I}^{-1}=6.4\times 10^9$.)  Hence, if the effects of the gravitational
radiation 
reaction were accounted for, as other authors have done
(e.g.{}, Oohara \& Nakamura 1990; Davies et al.\ 1994;
Zhuge, Centrella, \& McMillan 1994; Ruffert, Janka, \&
Sch\"afer 1996), the merger
of the $n=1.0$  binary would proceed on a timescale comparable to
the dynamical timescale.

The variation between the results of our
stability analyses of binaries with the softer equations of state
and those of RS (1992, 1995)
emphasizes the importance of comparing 
the results of hydrodynamic simulations performed with different
numerical techniques.
Certainly
the inclusion of more complex physics, such as the effects of the
gravitational radiation reaction and of full general relativity
should be the aim of research in this field.
However, an effort should be made to reach agreement on
the results of much simpler, Newtonian simulations in order that
their results and the results of more complex
simulations can be confidently presented as guides to 
those who will be designing and building gravitational radiation detectors
and interpreting the data collected at future gravitational
radiation observatories.

%
\acknowledgments
This work has been supported, in part, by funding from the National Science
Foundation through grants AST-9008166, AST-9528424, and PHY-9208914,
from NASA through
grants NAGW-2447 and NAG5-2777,
from the Louisiana State Board of
Regents through grant LEQSF (1995-1998)-115-35-4412, and from
the Baton Rouge, Louisiana Branch of the American Association of University
Women through a Doctoral Candidate Scholarship and by computational 
resources
from the Pittsburgh Supercomputing Center through grant PHY910018P.
The great majority of the numerical simulations presented herein were
carried out on computers in the Department of Physics and Astronomy's
Concurrent Computing Laboratory for Materials Simulation which
was established through grants LEQSF(1990-92)-ENH-12 and
LEQSF(1993-94)-ENH-TR05
from the Louisiana State Board of Regents.
Several of the numerical simulations also were carried out at the
Scalable Computing Laboratory of the
DOE Ames Laboratory at Iowa State University.
We would also like to thank Joan Centrella, Dimitris Christodoulou,
Greg Cook, Dong Lai, Dustin Laurence, Kip Thorne, and Horst V\"ath
for useful conversations and assistance.

\appendix

\section{Equilibrium Sequences}

Eleven WD equation of state equilibrium binary sequences,
each with a different $M_{T}$, are displayed in Figure 16.
See \S 3.2 for details.

\clearpage
%

\newpage
\centerline{\bf{FIGURE CAPTIONS}}

Fig. 1.--Polytropic Equilibrium Sequences.
Sequences of binaries
with polytropic indices $n=0.5$, 1.0, and 1.5 are displayed.
Each sequence displays the total energy $E$ and the total
angular momentum $J$ of synchronized equilibrium binaries with the same total mass
$M_{T}$ but changing separation $a$.  See the text (\S3.1)
for details on how to scale
the $M_{T}$ of each polytropic sequence.  Quantities have been normalized to
$G$, $M=M_{T}/2$, and $R_{Sph}$, where $R_{Sph}$ is the radius of a spherical
polytrope of mass $M$ and polytropic index $n$.  Points M mark binary systems
with the minimum $E$ and $J$ along each sequence.
Points C mark the system where
the surfaces of the stars come into contact. Points T mark the terminal model.

Fig. 2.--Images of a Detached Binary and a Dumbbell.
Example isodensity images of a detached binary and a contact binary,
or ``dumbbell.''  These two binaries have $n=1.0$; the separation of
the detached binary shown in Figure 2a is $a=3.28$; the separation of
the dumbell shown in Figure 2b is $a=2.70$.
The density level
is $5.0\times 10^{-3}$ of the maximum density.

Fig. 3.--Comparison with Hachisu's Polytropic Sequences.
The square of the angular velocity $\Omega_0$ versus
the square of the total angular momentum $J$ of equilibrium binaries with the
same total mass $M_T$ but decreasing separation $a$ is shown for both the
$n=0.5$ and 1.5
sequences of Hachisu (1986b, individual models connected by solid lines)
and for our sequences (individual models marked with +'s).  The quantities
are normalized to $4\pi G$, $M_T$, and $V_T$, the total volume of the
system.

Fig. 4.--White Dwarf Equilibrium Sequences.
The same as Figure 1 but for
binaries with the zero--temperature white dwarf equation of state.
Because, unlike the polytropes, the $M_T$ of these systems
cannot be scaled, separate sequences must be constructed
for binaries with different $M_T$.  Four representative sequences with $M_T = .500$,
1.19, 2.03, and 2.72 $M_\odot$ are shown here.  In Appendix A, sequences for
seven other values
of $M_T$ are shown, along with those given here.  The normalization
is the same as in Figure 1, but here $R_{Sph}$ is the radius of a spherical
model with the WD equation of state and mass $M=M_T/2$.

Fig. 5.--Separation of Models on White Dwarf Equilibrium Sequences.
The separation $a$ of the models at the points of contact ($\times$), the
minima (asterisks), and the terminal points (plusses) on the white dwarf
equilibrium sequences are shown as a function of the total mass $M_T$ of
the binaries on those sequences.

Fig. 6.--Comparison with Hachisu's White Dwarf Sequences.
The total
angular momentum $J$ versus the total mass $M_T$ of the models with the 
minimum $J$ on each of Hachisu's (1986b) constant maximum density
sequences (individual models marked with $\times$'s) and on each of our
constant $M_T$ sequences (individual models marked with +'s).
The angular momentum has been normalized to $10^{50}$ in cgs units
and the total mass to $M_{\odot}$.

Fig. 7.--Neutron Star Equilibrium Sequences.
The same as Figure 1
but for binaries with the F, FPS, and L realistic neutron star equations
of state.  The $M_T$ on each of these sequences is 2.80 $M_{\odot}$ (unlike
polytropes, the $M_T$ of these systems cannot be scaled).  The normalization
is the same as in Figure 1 but here $R_{Sph}$ is the radius of a spherical
model constructed with the particular NS equation of state of the sequence
and a mass $M=M_{T}/2$.
 
Fig. 8.--Comparison of Stability Tests Performed with Different
Numerical Techniques.
The moment of inertia $I$, normalized to $(MR_{Sph}^{2})$,
as a function of time $t$, normalized to the initial orbital
period $P_I$, is given for dynamical stability tests of a binary with
the zero--temperature white dwarf equation of state, $M_T=0.500\:M_{\odot}$,
and $a=2.63$.  The solid curve shows the results of a simulation performed with
our second--order accurate FDH code in the rotating frame; the dashed curve
shows the test performed with the same code but in the inertial frame; and
the dot--dashed curve shows the test performed in the rotating frame, but with
the advection terms in the FDH code reduced to first--order accuracy.

Fig. 9.--Stability Tests of the $M_T = 2.03\:M_{\odot}$ White Dwarf Sequence.
The variables in the lower plot are the same as Figure 8, but the simulations all are
performed in the rotating reference frame with the second--order
accurate FDH code.  The solid curve is the stability test of the terminal
model on the sequence ($a=2.53$); the dashed curve is the test of the model with
$a=2.63$; the dot--dashed curve is the test of the model at the minimum
($a=2.67$); and the
triple dot--dashed curve is the test of the model just past point of contact
($a=2.80$).  The moments of inertia of the binary models at the points of
contact, minima, and termination along the equilibrium sequence are
labelled with the letters C, M, and T, respectively.  For convenience,
the sequence itself has been reprinted from Figure 4 in the upper plot.

Fig. 10.--Evolution of Stable Binary.
Isodensity images of a stable binary with the WD equation of state,
$M_T=2.03\:M_{\odot}$,
and $a=2.67$.  The density level
is $5\times 10^{-4}$ of the maximum density.
Figure 10a was taken at 
$t=0.0\:P_{I}$; 10b at $t=0.8\:P_{I}$; 10c at $t=1.5\:P_{I}$;
10d at $t=2.3\:P_{I}$; 10e at $t=3.0\:P_{I}$; and 10f at
$t=3.8\:P_{I}$.
Since the simulation
was performed in the rotating frame, the binary does not appear to pivot
very much in this set of images.

Fig. 11.--Stability Tests of the $M_T = .500\:M_{\odot}$
White Dwarf and the $n=1.5$ Sequences.
The same variables
and code as used in Figure 9.  The solid curve is the simulation of the
terminal model ($a=2.45$) on the $M_T = .500\:M_{\odot}$ white
dwarf sequence (the equilibrium sequence itself is reprinted from Figure 4
in the upper
plot); the dashed curve is the test of the model on this same
sequence with $a=2.63$; and the dot--dashed curve is the test of the
terminal model ($a=2.45$) on the $n=1.5$ polytropic sequence.

Fig. 12.--Stability Tests of $n=1.0$ Sequence.
The same variables
and code as described in Figure 9.  The solid curve is the test of the
terminal model ($a=2.62$); the dashed curve is the test of the model with
$a=2.70$; the dot--dashed curve is the test of the model at the minimum
of the sequence ($a=2.88$); and the triple dot--dashed curve is the
test of the model with $a=3.28$. The equilibrium sequence itself is
reprinted from Figure 1 in the upper plot.

Fig. 13.--Stability Tests of $n=0.5$ Sequence.
The same variables
and code as described in Figure 9.  The solid curve is the test of the
terminal model ($a=2.77$); the short dashed curve is the test of the model
at the minimum
of the sequence ($a=3.10$); the dot--dashed curve is the test of the model
with $a=3.17$; the triple dot--dashed curve is the
test of the model with $a=3.24$; and the long dashed curve is the test
of the model with $a=3.41$. The equilibrium sequence itself is
reprinted from Figure 1 in the upper plot.

Fig. 14.--Stability Tests of Equation of State L Sequence.
The same variables
and code as described in Figure 9.  The solid curve is the test of the model
near the minimum of the sequence ($a=2.79$); the dashed curve
is the test of the model with $a=3.26$.
The equilibrium sequence itself is reprinted from Figure 7 in the upper plot.

Fig. 15.--Stability Tests of Equation of State F Sequence.
The same variables
and code as described in Figure 9.  The solid curve is the test of the system
at the minimum of the sequence ($a=2.80$).  The equilibrium sequence itself is
reprinted from Figure 7 in the upper plot.

Fig. 16.--White Dwarf Equilibrium Sequences.
$M_{T}=.298$, $.500$, $.803$, $1.19$,
$1.63$, $2.03$, $2.36$, $2.58$,
$2.72$, $2.81$, and $2.85\:M_{\odot}$.


\newpage
\begin{table}
\clearpage
\caption{Lane-Emden Constants, $m_n$ and $r_n$\tablenotemark{a}}
\bigskip
\begin{tabular} {l l l}
$n$ & $m_n$ & $r_n$ \\
0.5 & 3.7871 & 2.7528 \\
1.0 & 3.14159 & 3.14159 \\
1.5 & 2.71406 & 3.65375 \\
\end{tabular}
\tablenotetext{a}{From Chandrasekhar (1967)}
\end{table}
\clearpage

\begin{table}
\clearpage
\caption{Separation of Models at Minima of Polytropic Sequences}
\bigskip
\begin{tabular} {cl l l}
 Technique, Authors & $n=0.5$ & $n=1.0$ & $n=1.5$ \\
 Analytic, LRS & 2.99{\tablenotemark{a}} & 2.76{\tablenotemark{a}}
& 2.55{\tablenotemark{a}} \\
 SPH, RS & --- & 2.90{\tablenotemark{a}} & 2.67{\tablenotemark{b}} \\
HSCF, New \& Tohline & 3.20{\tablenotemark{c}} & 2.98{\tablenotemark{c}} & 2.77{\tablenotemark{c}} \\
HSCF, New \& Tohline & 3.11{\tablenotemark{d}} & 2.89{\tablenotemark{d}} & 2.70{\tablenotemark{d}} \\
\end{tabular}
\tablenotetext{a}{From LRS (1993b), separation between components' centers of
mass}
\tablenotetext{b}{From RS (1995), separation between components' centers of mass}
\tablenotetext{c}{From this work, separation between components' centers of
mass}
\tablenotetext{d}{From this work, separation between components' pressure
maxima}
\end{table}
\clearpage

\begin{table}
\clearpage
\caption{Comparison of $64^3$ and $128^3$ Sequences}
\bigskip
\begin{tabular} {c c c c c}
 EOS, $M_T$ & Grid Size & $a(\rm{C})$ & $a(\rm{M})$ & $a(\rm{T})$ \\
 WD, $2.03\:M_{\odot}$ & $64^3$ & 2.84 & 2.67 & 2.53 \\
                       & $128^3$ & 2.84 & 2.68 & 2.53 \\
 WD, $.500\:M_{\odot}$ & $64^3$ & 2.81 & 2.70 & 2.45 \\
                       & $128^3$ & 2.81 & 2.70 & 2.46 \\
 $n=1.5$ & $64^3$ & 2.81 & 2.70 & 2.45 \\
         & $128^3$ & 2.81 & 2.70 & 2.45 \\
 $n=1.0$ & $64^3$ & 2.84 & 2.88 & 2.62 \\
         & $128^3$ & 2.84 & 2.89 & 2.61 \\
 $n=0.5$ & $64^3$ & 2.94 & 3.10 & 2.77 \\
         & $128^3$ & 2.94 & 3.11 & 2.76 \\
\end{tabular}
\end{table}

\clearpage
\begin{figure}
\plotone{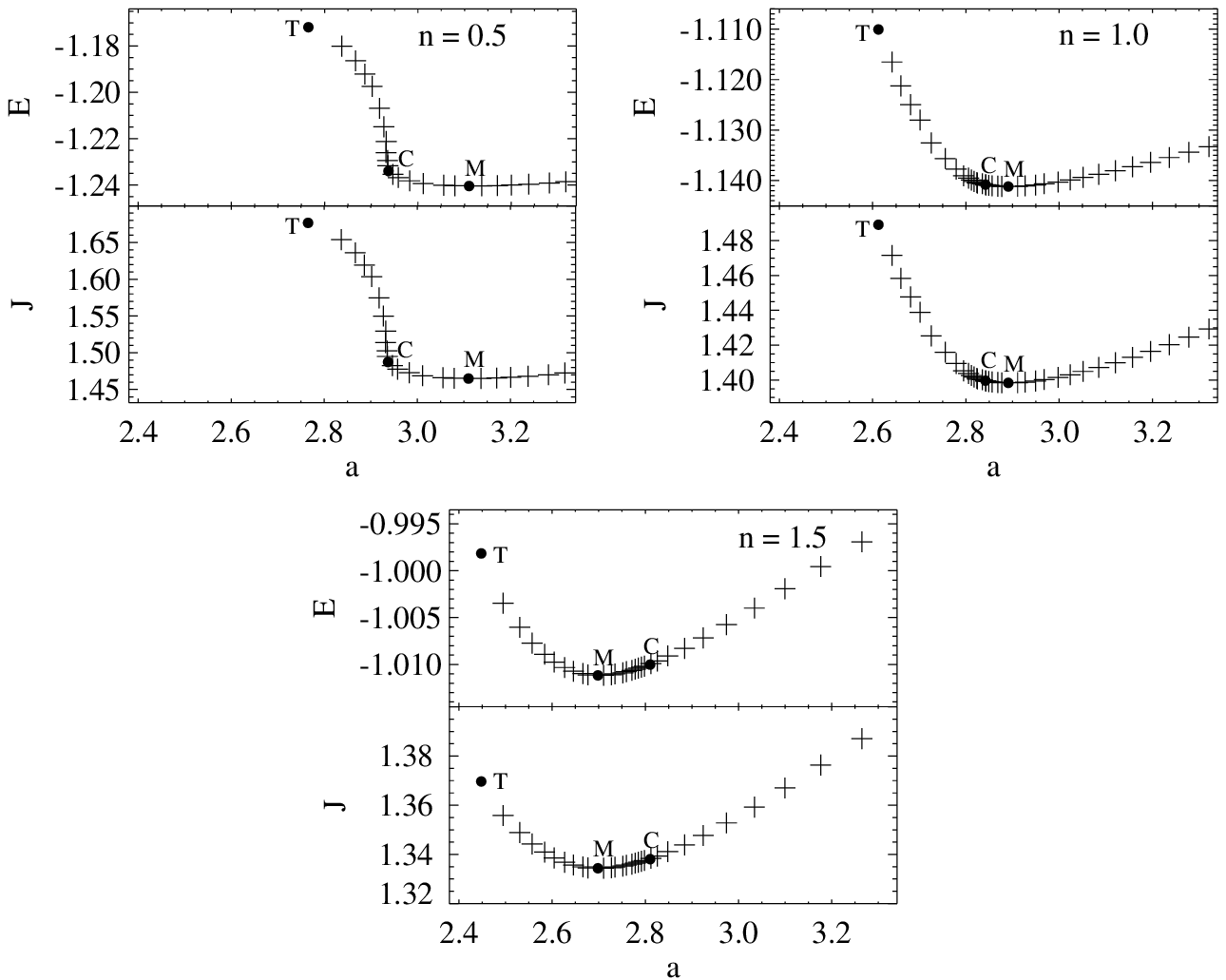}
\end{figure}
\clearpage

\clearpage
\begin{figure}
\plotone{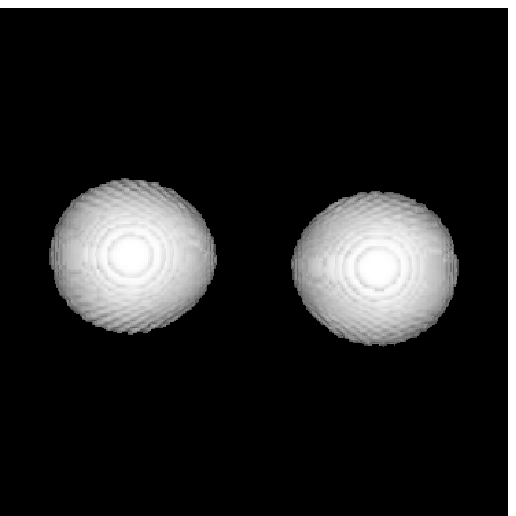}
\end{figure}
\clearpage

\clearpage
\begin{figure}
\plotone{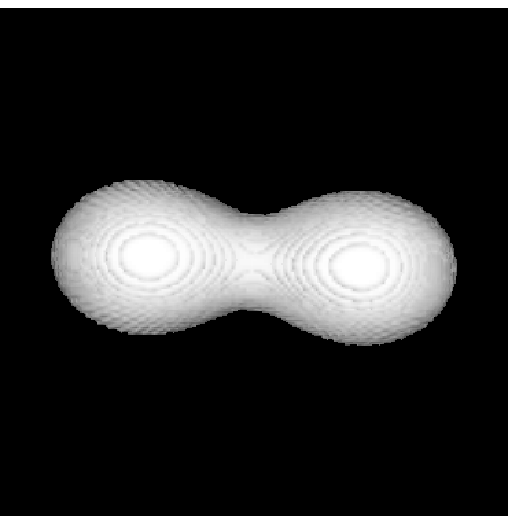}
\end{figure}
\clearpage

\clearpage
\begin{figure}
\plotone{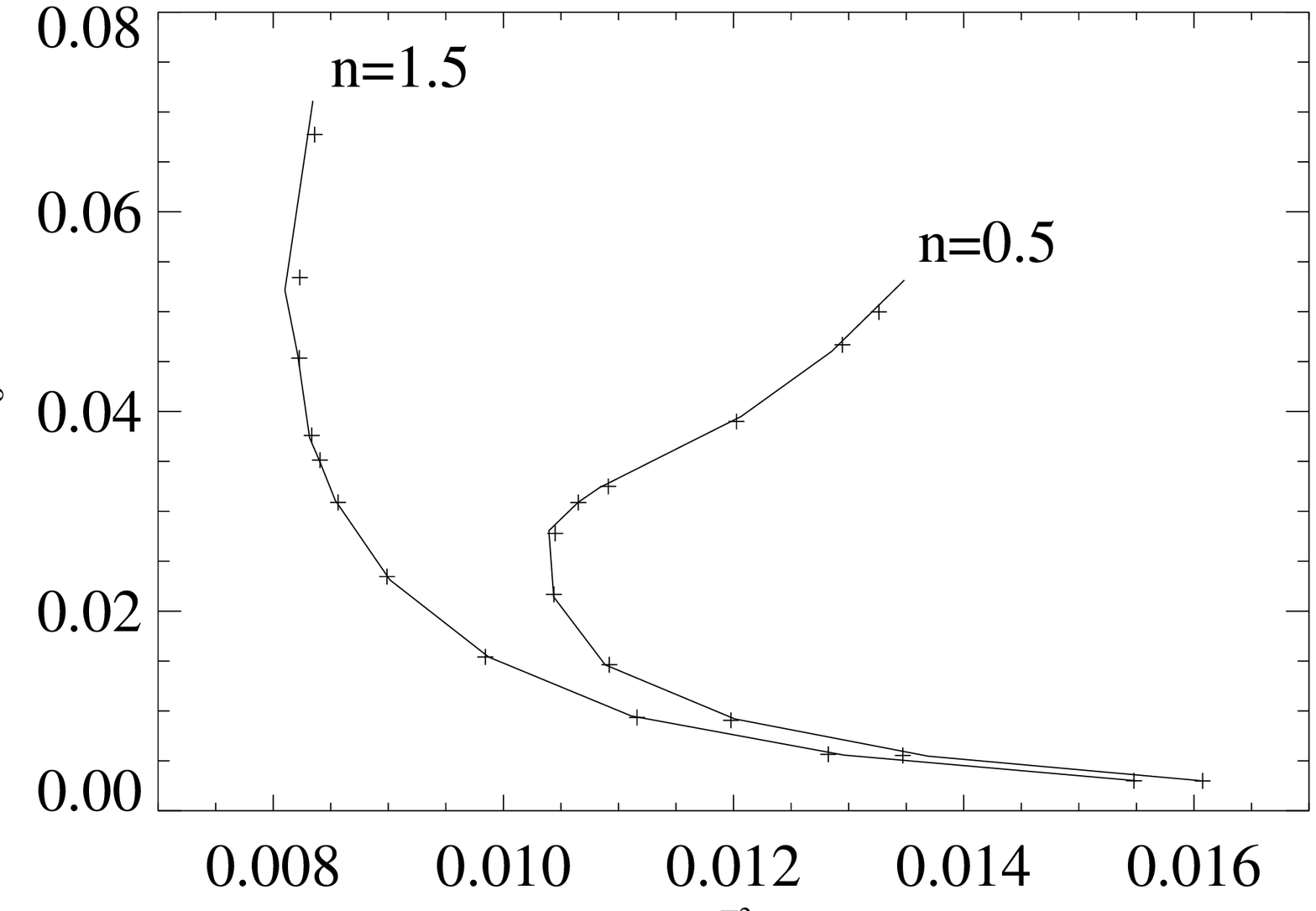}
\end{figure}
\clearpage

\clearpage
\begin{figure}
\plotone{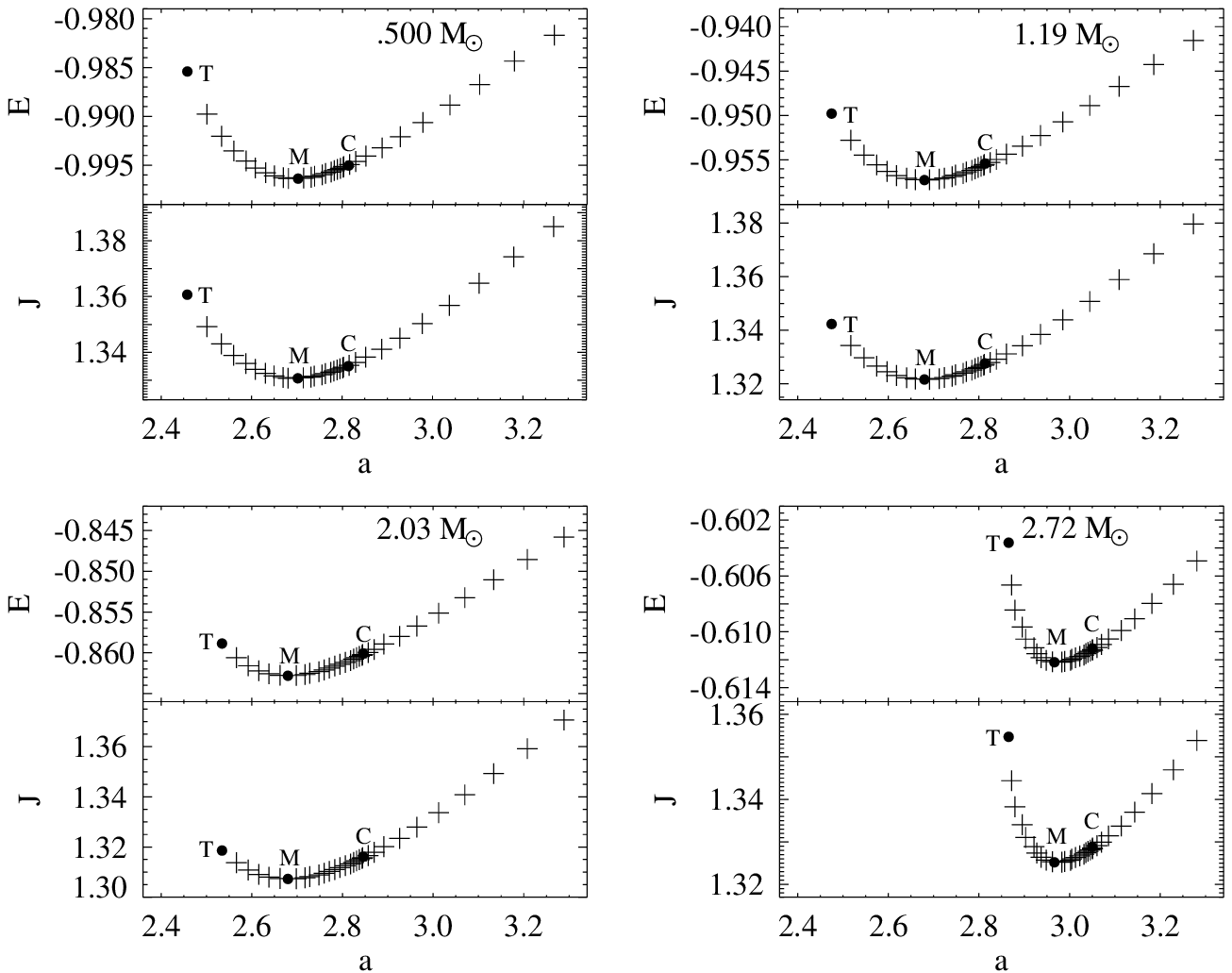}
\end{figure}
\clearpage

\clearpage
\begin{figure}
\plotone{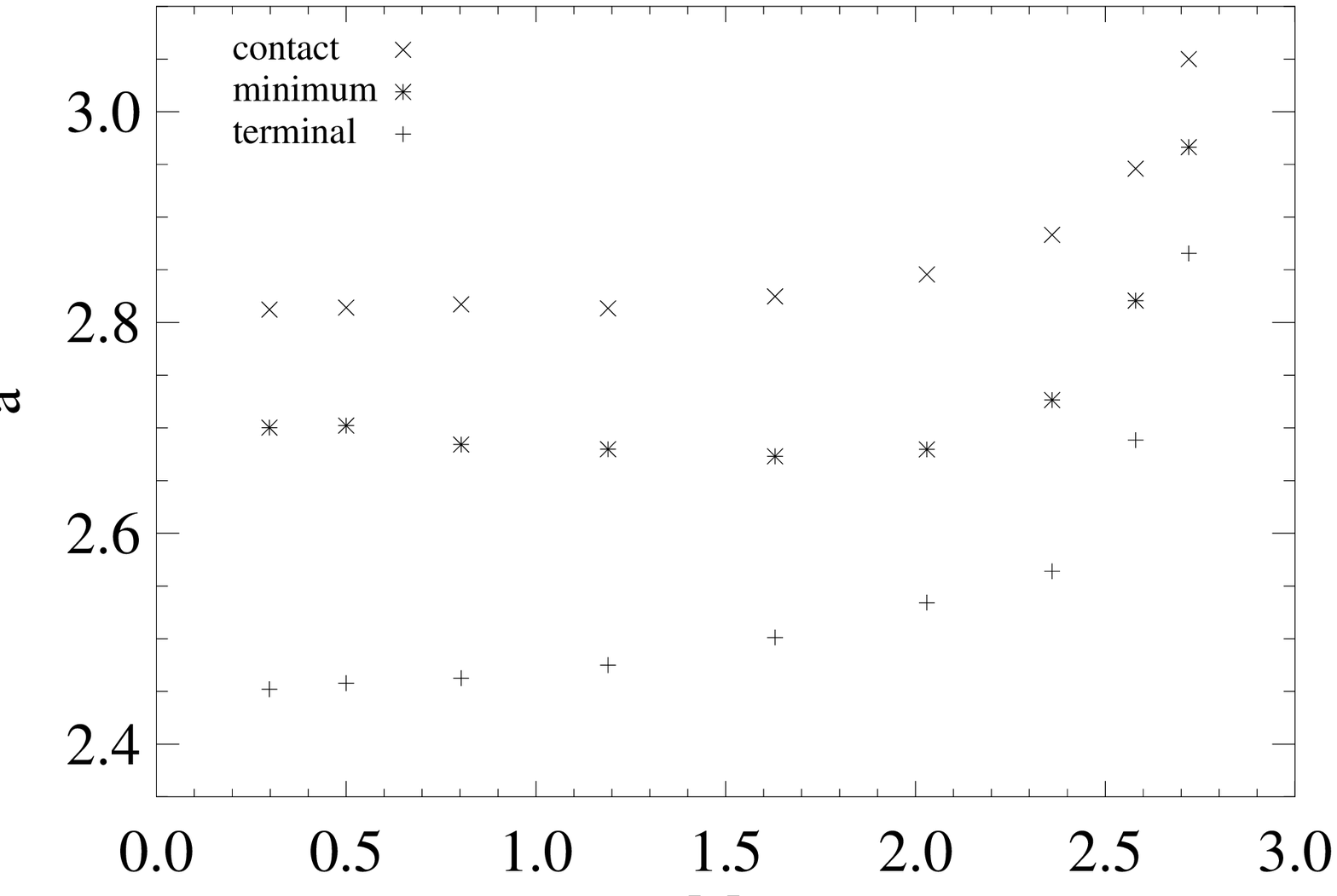}
\end{figure}
\clearpage

\clearpage
\begin{figure}
\plotone{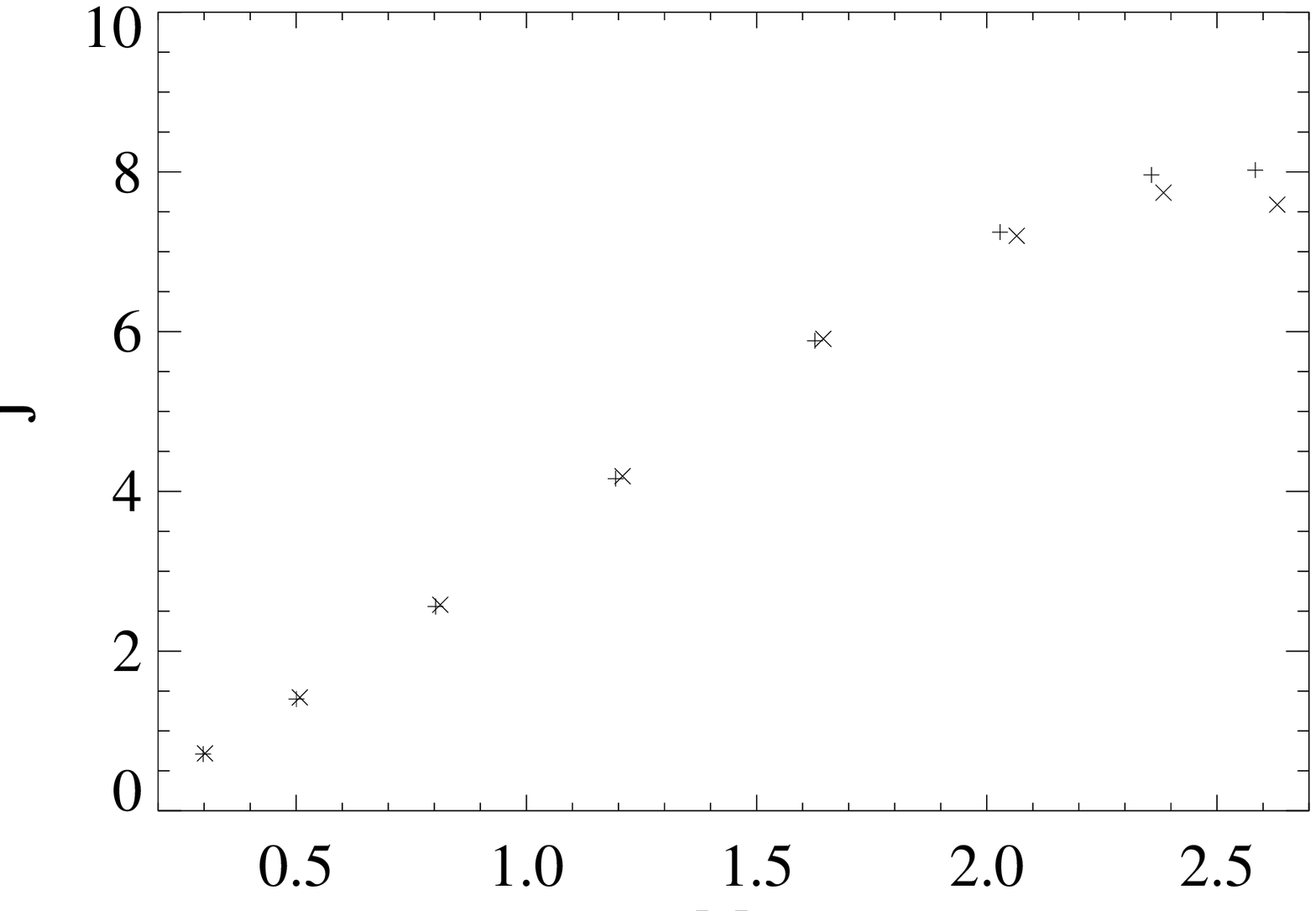}
\end{figure}
\clearpage

\clearpage
\begin{figure}
\plotone{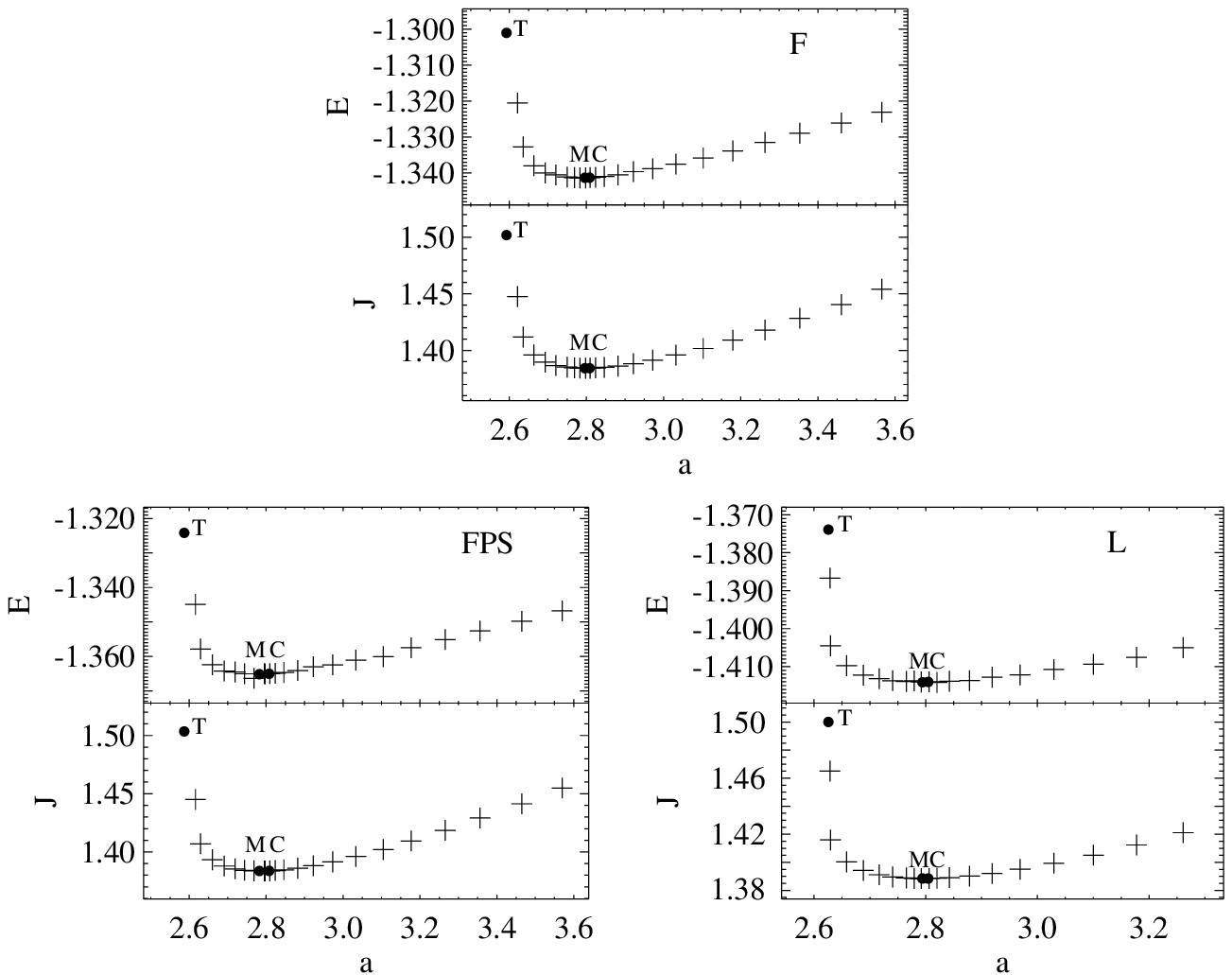}
\end{figure}
\clearpage

\clearpage
\begin{figure}
\plotone{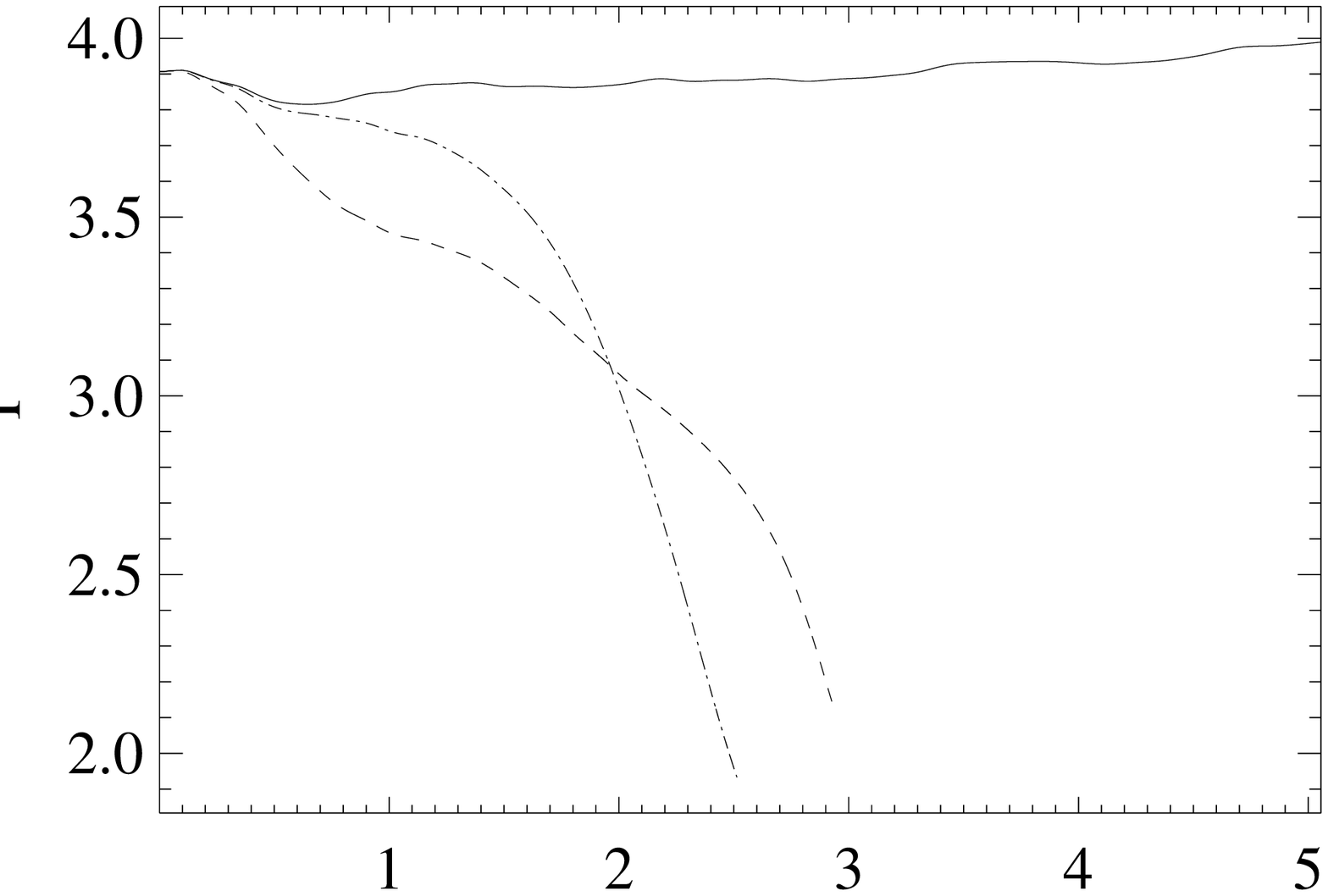}
\end{figure}
\clearpage

\clearpage
\begin{figure}
\plotone{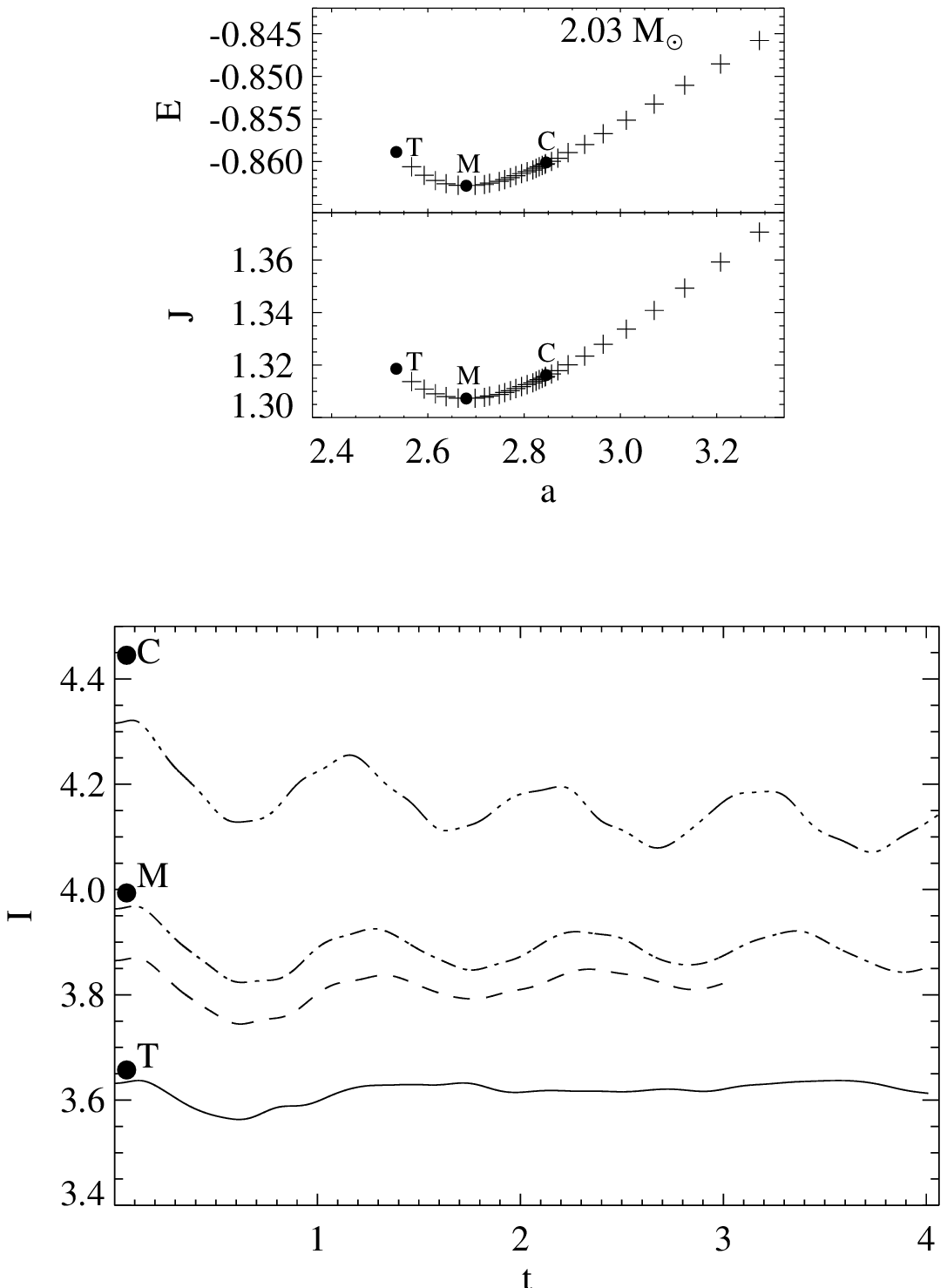}
\end{figure}
\clearpage

\clearpage
\begin{figure}
\plotone{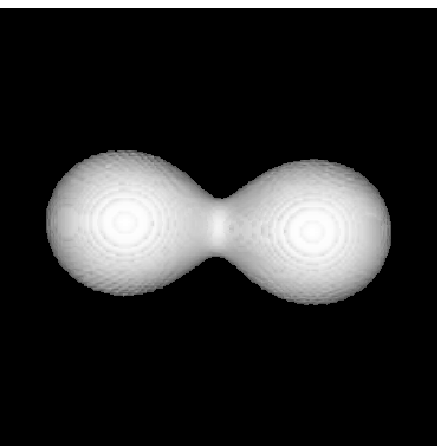}
\end{figure}
\clearpage

\clearpage
\begin{figure}
\plotone{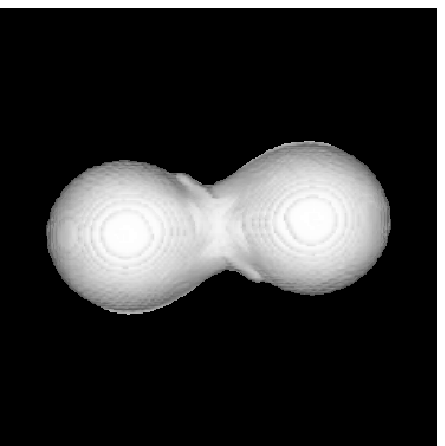}
\end{figure}
\clearpage

\clearpage
\begin{figure}
\plotone{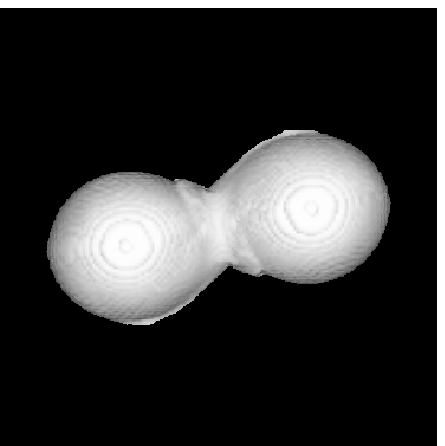}
\end{figure}
\clearpage

\clearpage
\begin{figure}
\plotone{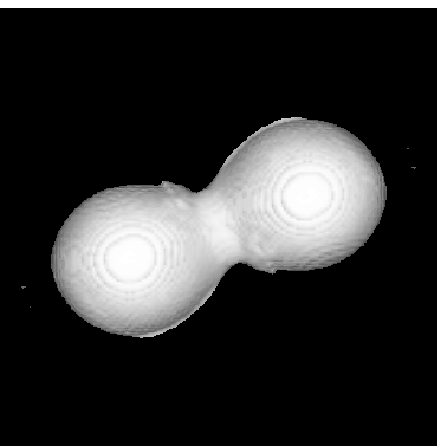}
\end{figure}
\clearpage

\clearpage
\begin{figure}
\plotone{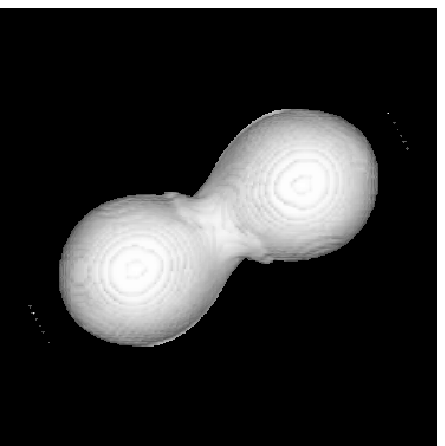}
\end{figure}
\clearpage

\clearpage
\begin{figure}
\plotone{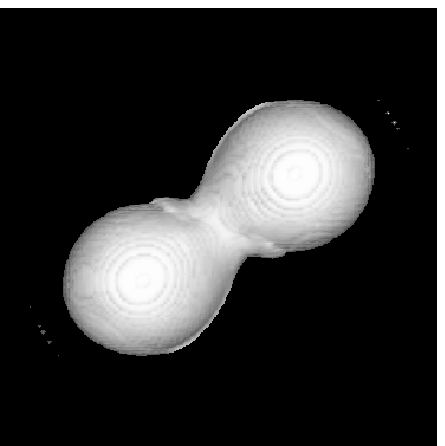}
\end{figure}
\clearpage

\clearpage
\begin{figure}
\plotone{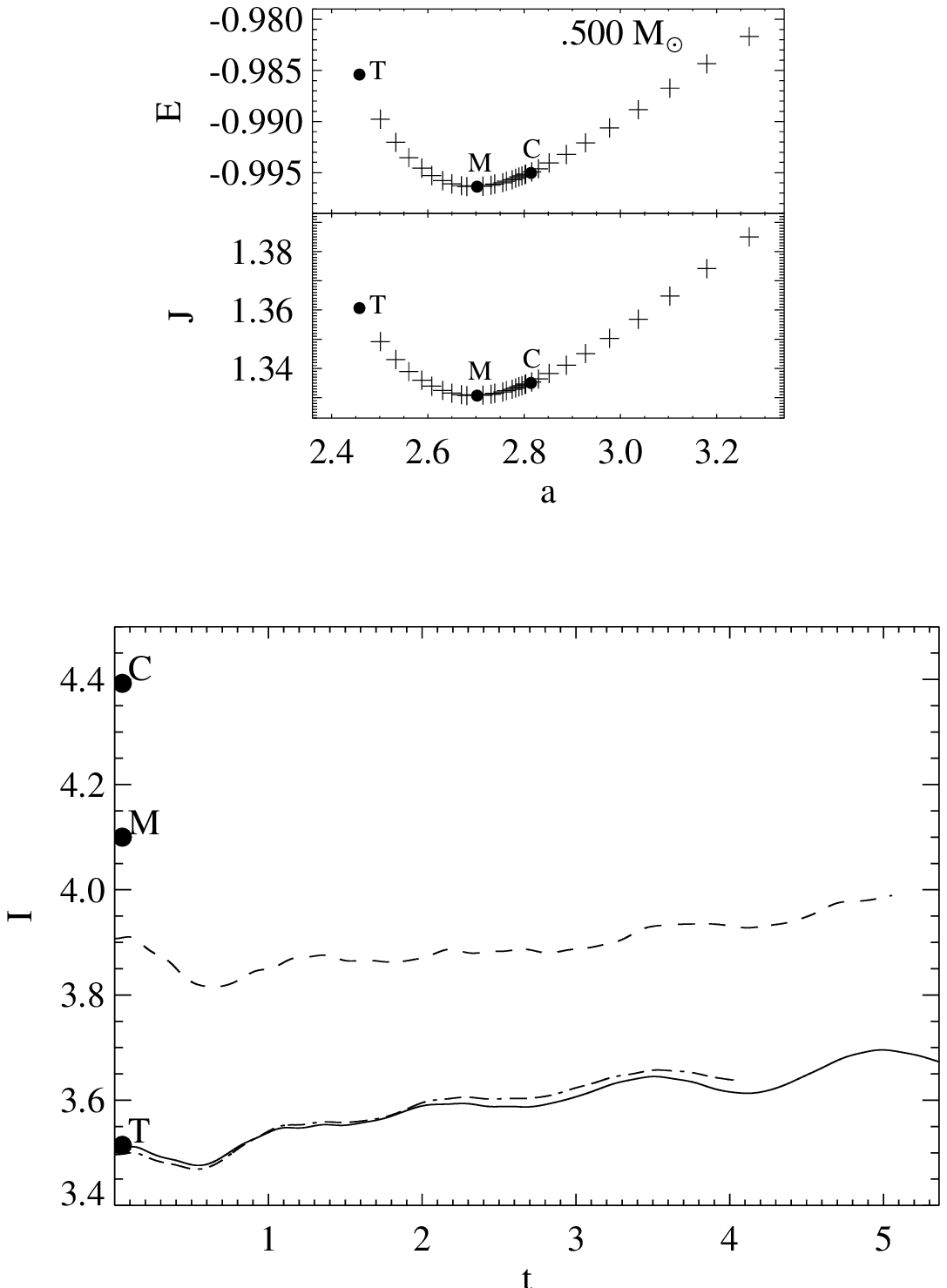}
\end{figure}
\clearpage

\clearpage
\begin{figure}
\plotone{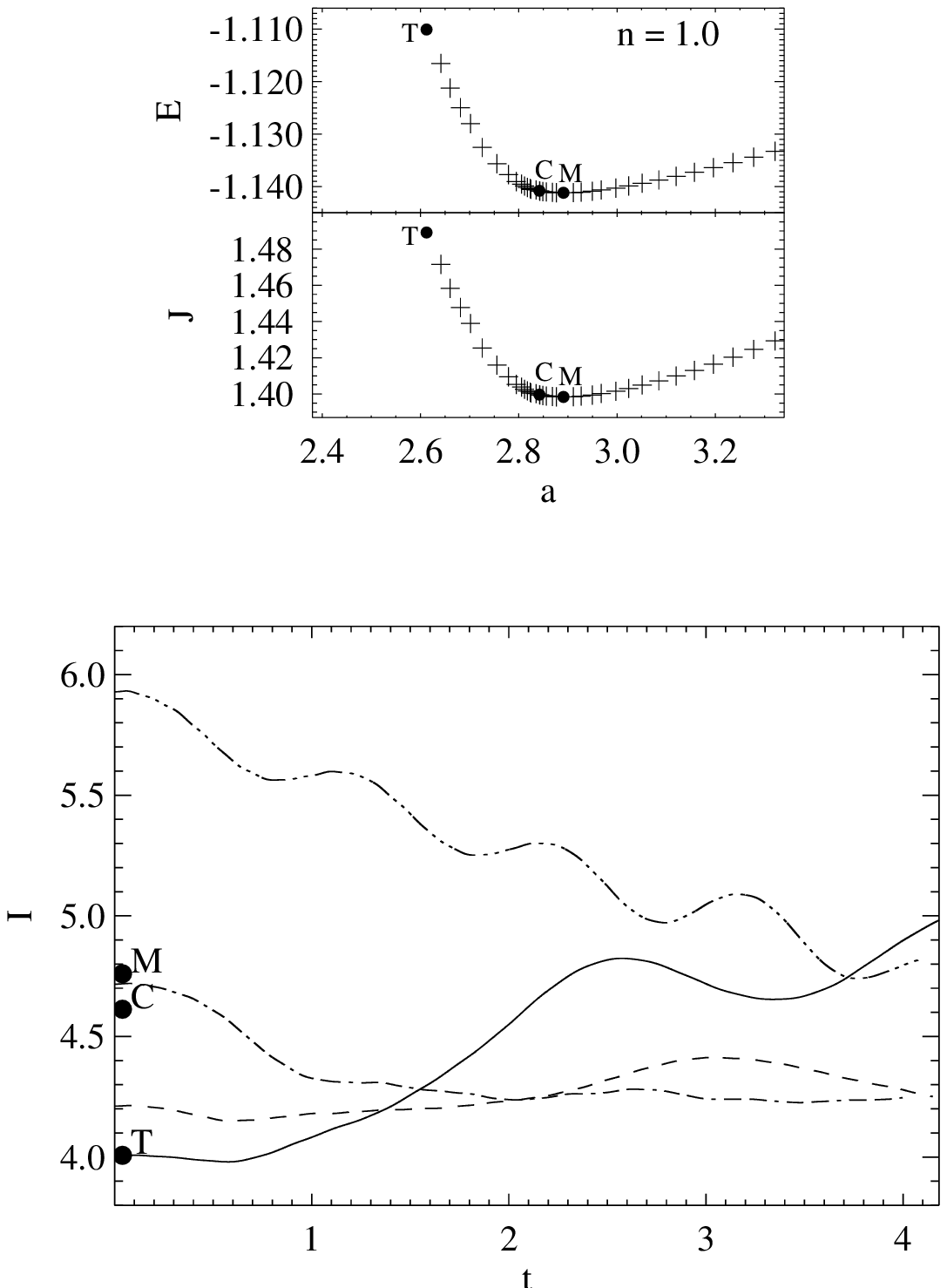}
\end{figure}
\clearpage

\clearpage
\begin{figure}
\plotone{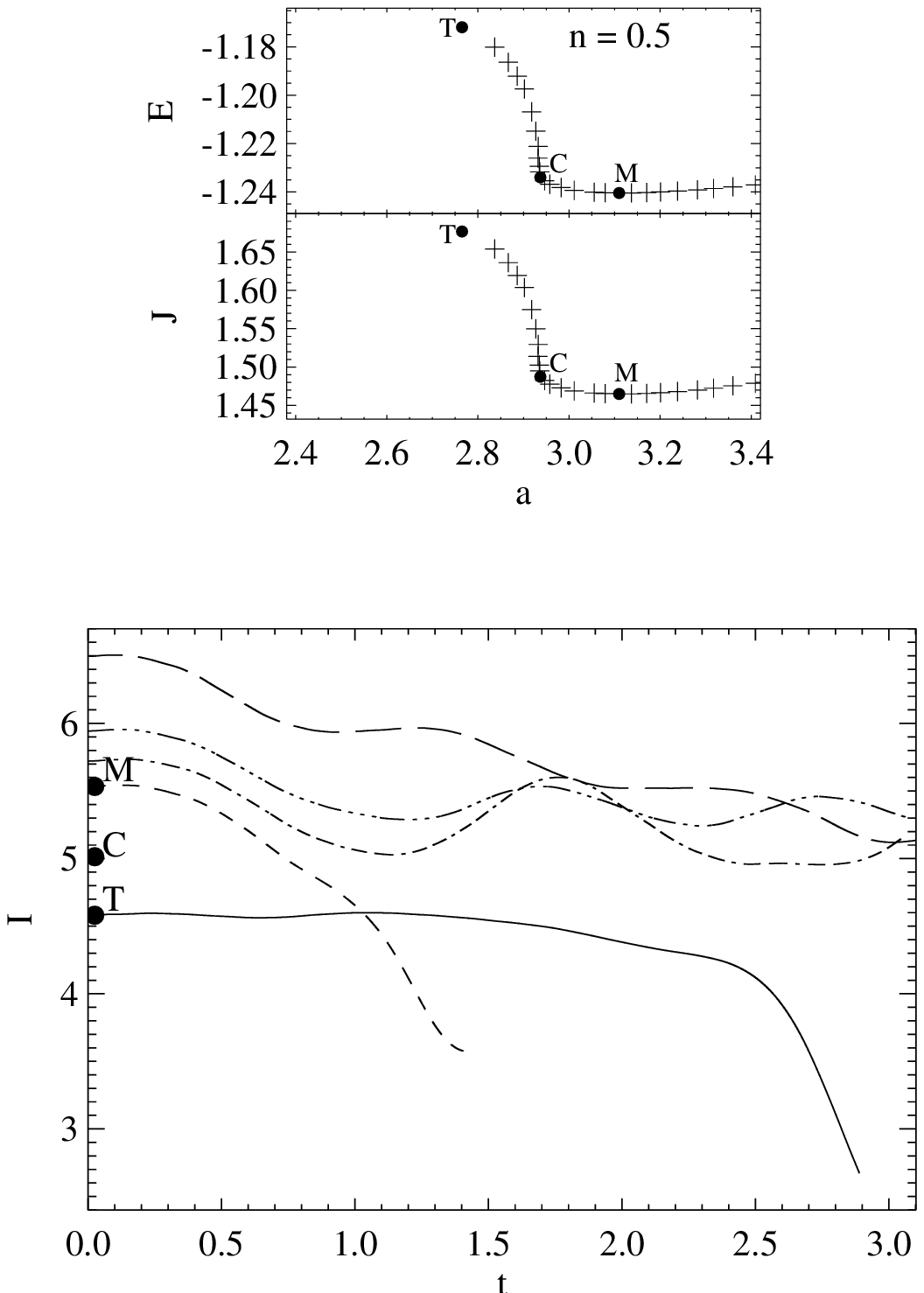}
\end{figure}
\clearpage

\clearpage
\begin{figure}
\plotone{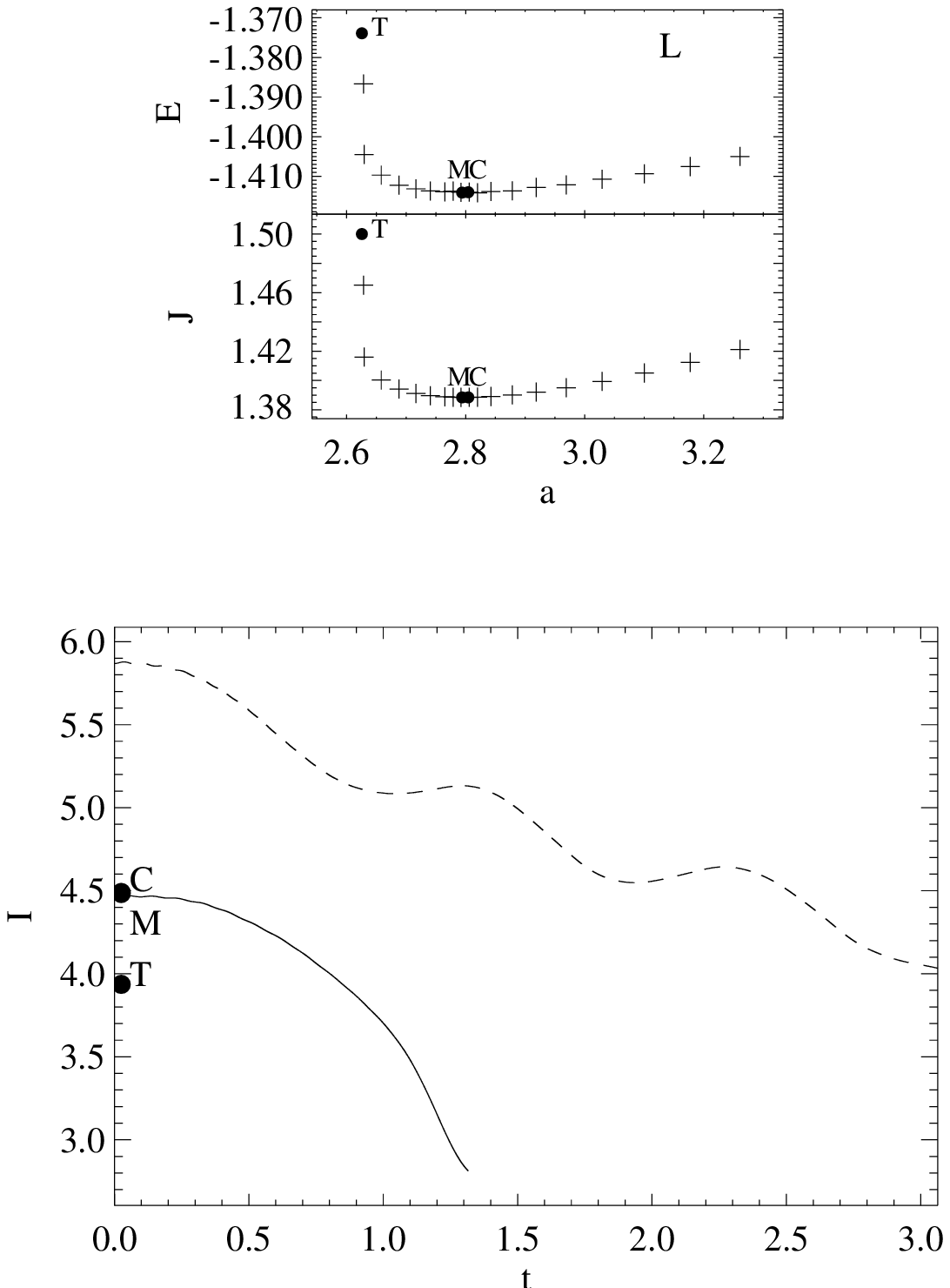}
\end{figure}
\clearpage

\clearpage
\begin{figure}
\plotone{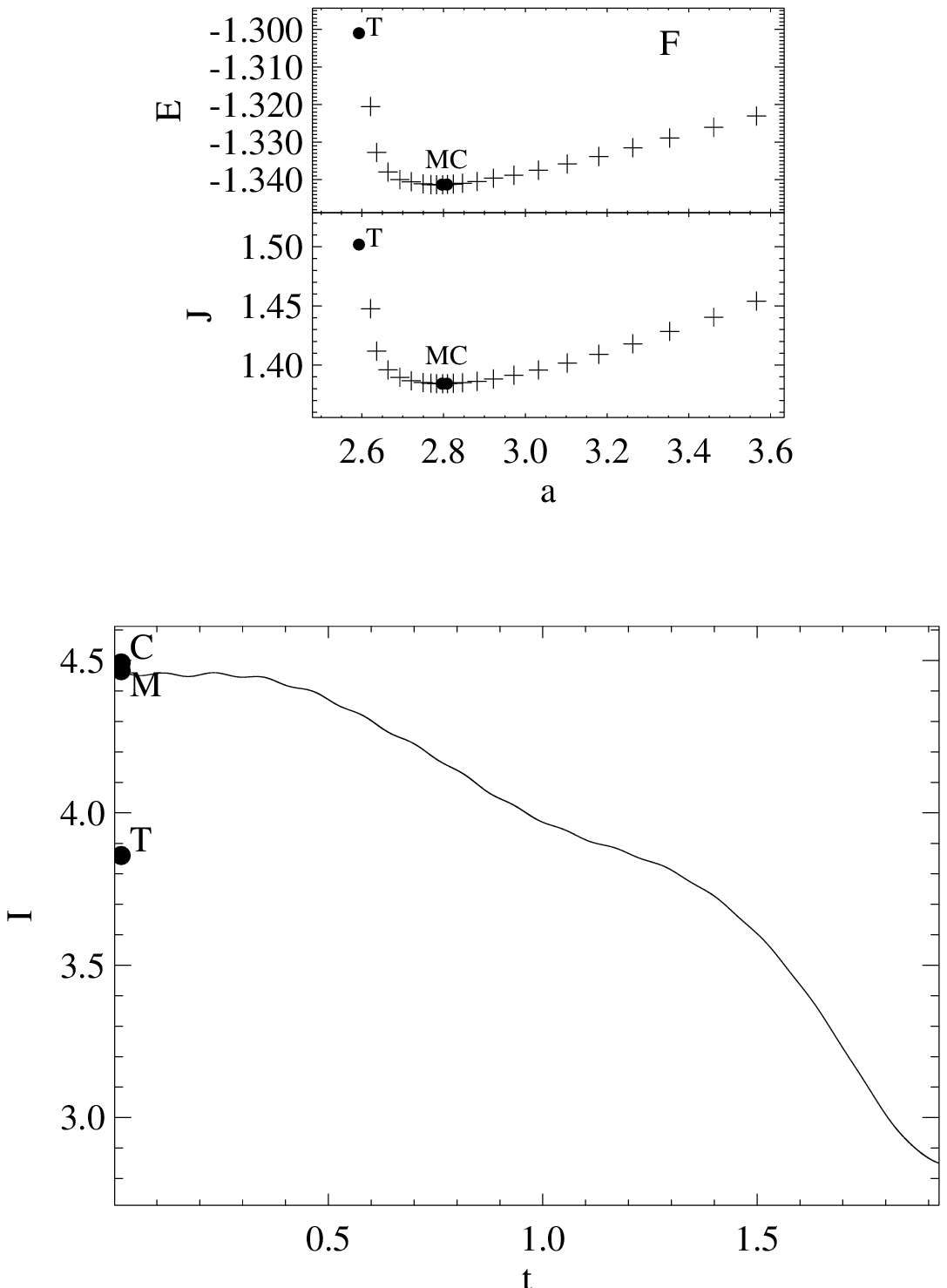}
\end{figure}
\clearpage

\clearpage
\begin{figure}
\plotone{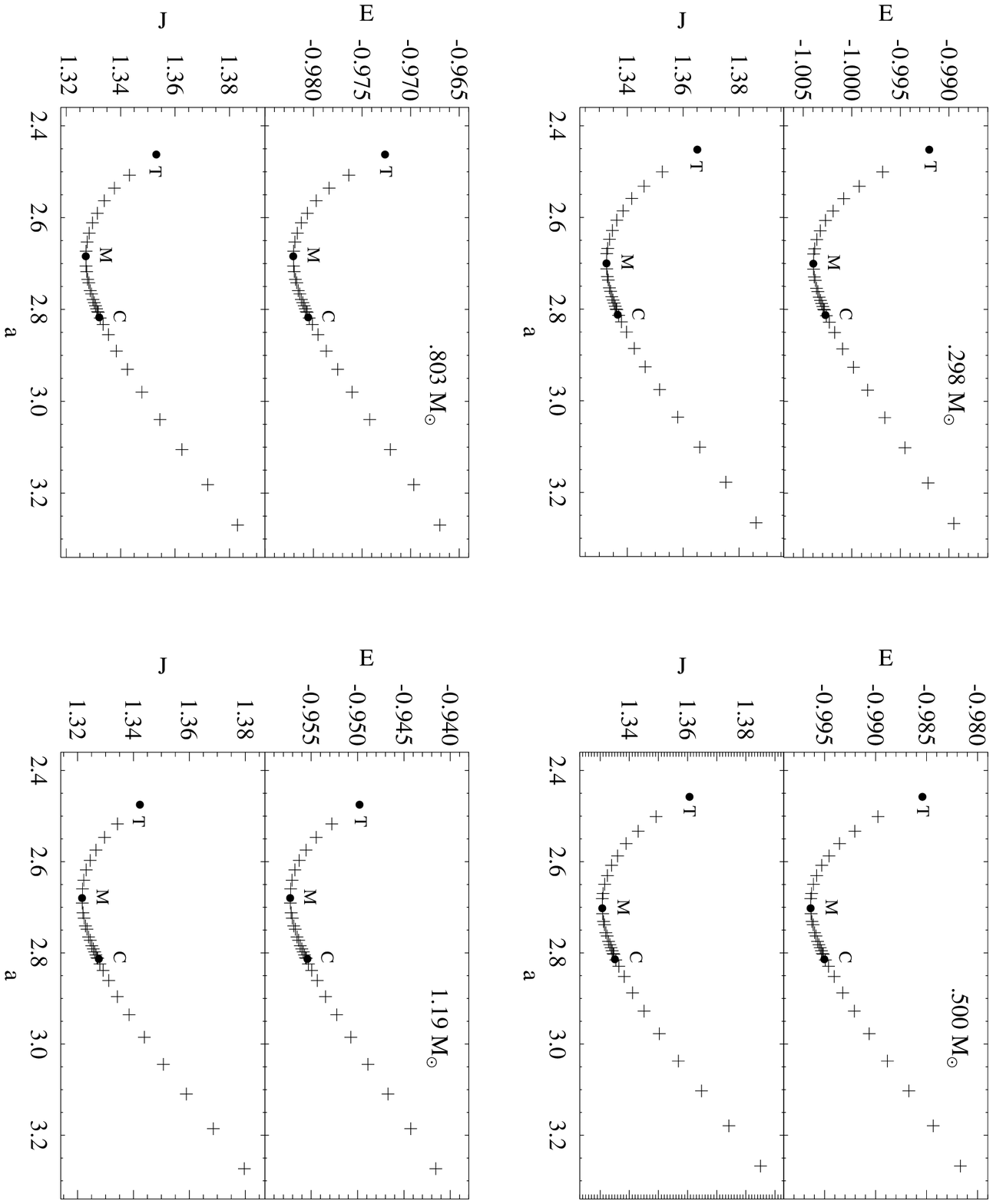}
\end{figure}
\clearpage

\clearpage
\begin{figure}
\plotone{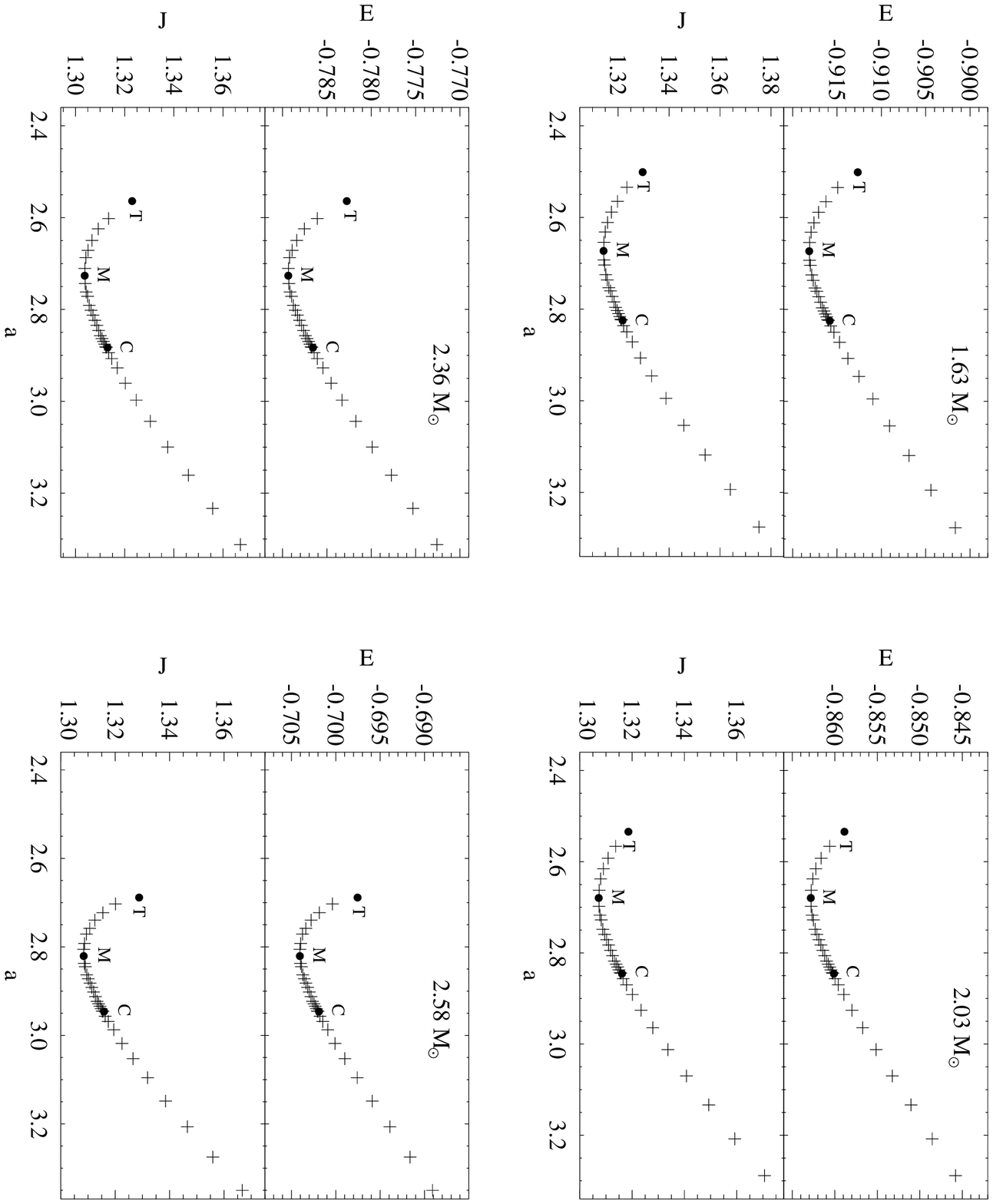}
\end{figure}
\clearpage

\clearpage
\begin{figure}
\plotone{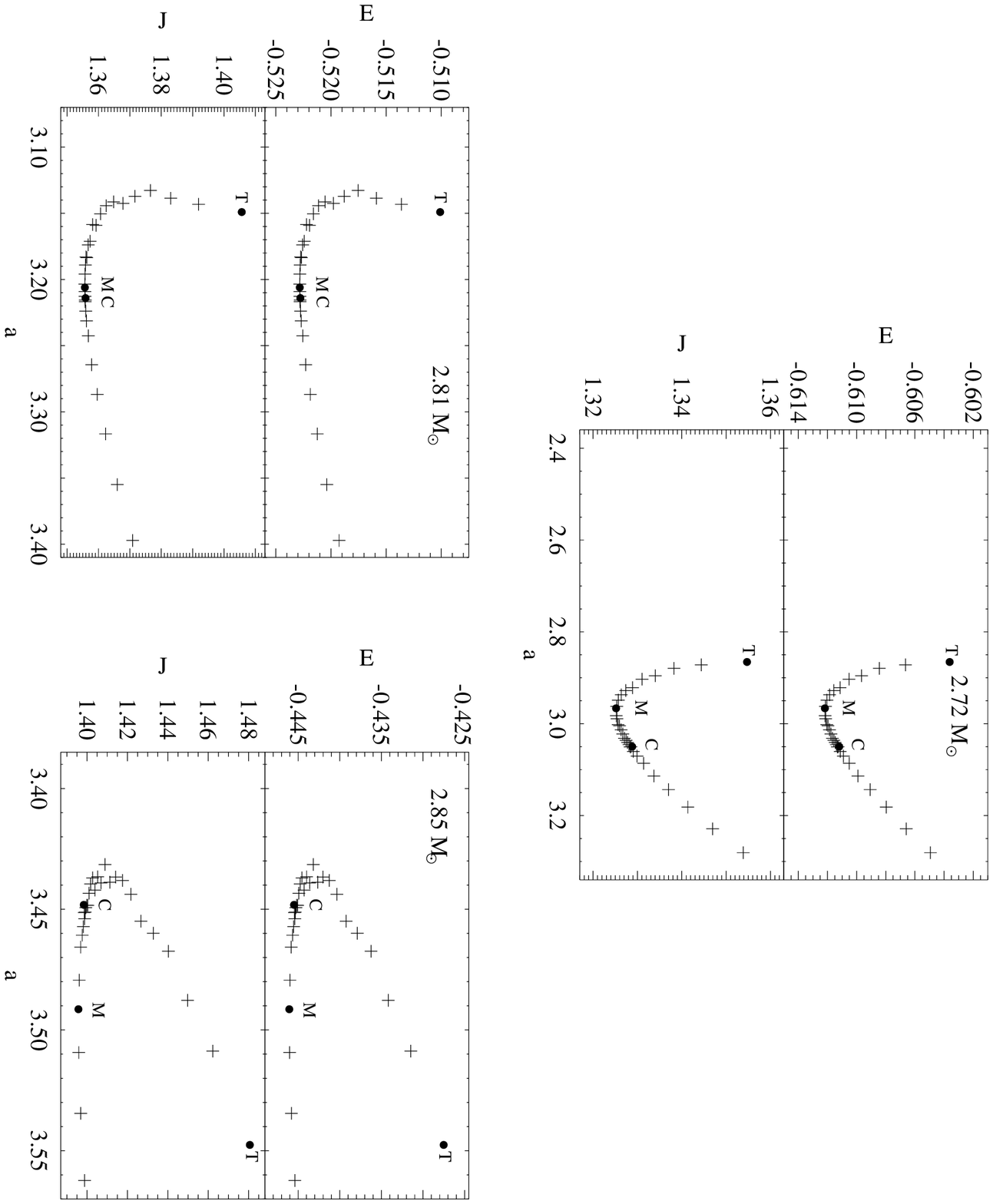}
\end{figure}
\clearpage

\end{document}